\documentclass[conference]{IEEEtran}
\usepackage[T1]{fontenc}
\usepackage[latin9]{inputenc}
\usepackage{float}
\usepackage{amsthm}
\usepackage{amsmath}
\usepackage{amssymb}
\usepackage{graphicx}
\usepackage{esint}

\makeatletter

\providecommand{\tabularnewline}{\\}
\floatstyle{ruled}
\newfloat{algorithm}{tbp}{loa}
\providecommand{\algorithmname}{Algorithm}
\floatname{algorithm}{\protect\algorithmname}

\theoremstyle{plain}
\newtheorem{thm}{\protect\theoremname}
\theoremstyle{definition}
\newtheorem{defn}[thm]{\protect\definitionname}
\theoremstyle{plain}
\newtheorem{lem}[thm]{\protect\lemmaname}
\theoremstyle{plain}
\newtheorem{cor}[thm]{\protect\corollaryname}

\usepackage{amsthm}

\providecommand{\tabularnewline}{\\}

\ifCLASSINFOpdf
\else
\fi

\makeatother

\providecommand{\corollaryname}{Corollary}
\providecommand{\definitionname}{Definition}
\providecommand{\lemmaname}{Lemma}
\providecommand{\theoremname}{Theorem}

\begin{document}

\title{{\Huge Probe and Adapt: Rate Adaptation for HTTP Video Streaming
At Scale}}


\author{\IEEEauthorblockN{Zhi Li, Xiaoqing Zhu, Josh Gahm, Rong Pan, Hao
Hu, Ali C. Begen, Dave Oran} \IEEEauthorblockA{Cisco Systems, San
Jose, CA USA\\
 \{zhil2, xiaoqzhu, jgahm, ropan, hahu2, abegen, oran\}@cisco.com}}




\maketitle

\begin{abstract}
Today, the technology for video streaming over the Internet is converging
towards a paradigm named HTTP-based adaptive streaming (HAS), which
brings two new features. First, by using HTTP/TCP, it leverages network-friendly
TCP to achieve both firewall/NAT traversal and bandwidth sharing.
Second, by pre-encoding and storing the video in a number of discrete
rate levels, it introduces video bitrate adaptivity in a scalable
way so that the video encoding is excluded from the closed-loop adaptation.
A conventional wisdom is that the TCP throughput observed by an HAS
client indicates the available network bandwidth, and thus can be
used as a reliable reference for video bitrate selection.

We argue that this is no longer true when HAS becomes a substantial
fraction of the total traffic. We show that when multiple HAS clients
compete at a network bottleneck, the presence of competing clients
and the discrete nature of the video bitrates together result in difficulty
for a client to correctly perceive its fair-share bandwidth. Through
analysis and test bed experiments, we demonstrate that this fundamental
limitation leads to, for example, video bitrate oscillation that negatively
impacts the video viewing experience. We therefore argue that it is
necessary to design at the application layer using a ``probe-and-adapt''
principle for HAS video bitrate adaptation, which is akin to, but
also independent of the transport-layer TCP congestion control. We
present PANDA -- a client-side rate adaptation algorithm for HAS --
as practical embodiment of this principle. Our test bed results show
that compared to conventional algorithms, PANDA is able to reduce
the instability of video bitrate selection by over 75\% without increasing
the risk of buffer underrun.
\end{abstract}


\IEEEpeerreviewmaketitle

\section{Introduction}

Over the past few years, we have witnessed a major technology convergence
for Internet video streaming towards a new paradigm named HTTP-based
adaptive streaming (HAS). Since its inception in 2007 by Move Networks
\cite{move07}, HAS has been quickly adopted by major vendors and
service providers. Today, HAS is employed for over-the-top video delivery
by many major media content providers. A recent report by Cisco \cite{CiscoWhitePaper}
predicts that video will constitute more than 90\% of the total Internet
traffic by 2014. Therefore, HAS may become a predominant form of Internet
traffic in just a few years.

In contrast to conventional RTP/UDP-based video streaming, HAS uses
HTTP/TCP -- the protocol stack traditionally used for Web traffic.
In HAS, a video stream is chopped into short segments of a few seconds
each. Each segment is pre-encoded and stored at a server in a number
of versions, each with a distinct video bitrate, resolution and/or
quality. After obtaining a manifest file with necessary information,
a client downloads the segments sequentially using plain HTTP GETs,
estimates the network conditions, and selects the video bitrate of
the next segment on-the-fly. A conventional wisdom is that since the
bandwidth sharing of HAS is dictated by TCP, the problem of video
bitrate selection can be resolved straightforwardly. A simple rule
of thumb is to approximately match the video bitrate to the observed
TCP throughput.

\begin{figure}
\begin{centering}
\hspace{-0.3in} %
\begin{minipage}[t]{1\columnwidth}%
\begin{center}
\hspace{0.2in}\footnotesize Fetched Bitrate Aggregated over 36 Streams\vspace{-0.23in}

\par\end{center}

\begin{center}
\includegraphics[scale=0.35]{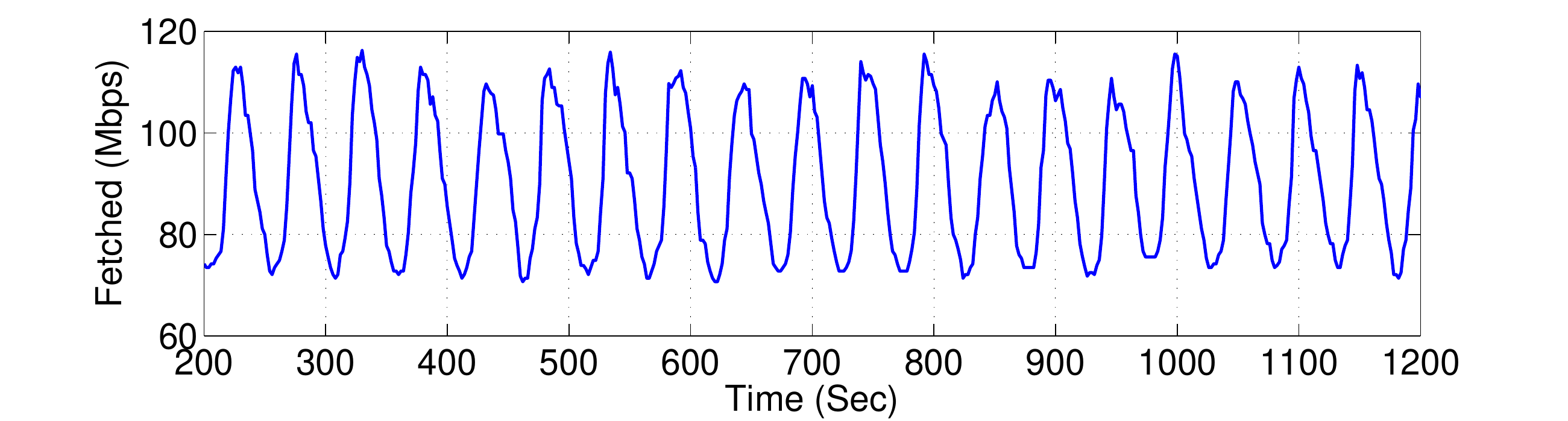}
\par\end{center}%
\end{minipage}
\par\end{centering}

\begin{centering}
\vspace{0.05in}
\hspace{-0.3in} %
\begin{minipage}[t]{1\columnwidth}%
\begin{center}
\hspace{0.2in}\footnotesize Fetched Bitrate of Individual Streams
(Zoom In)\vspace{-0.23in}

\par\end{center}

\begin{center}
\includegraphics[scale=0.35]{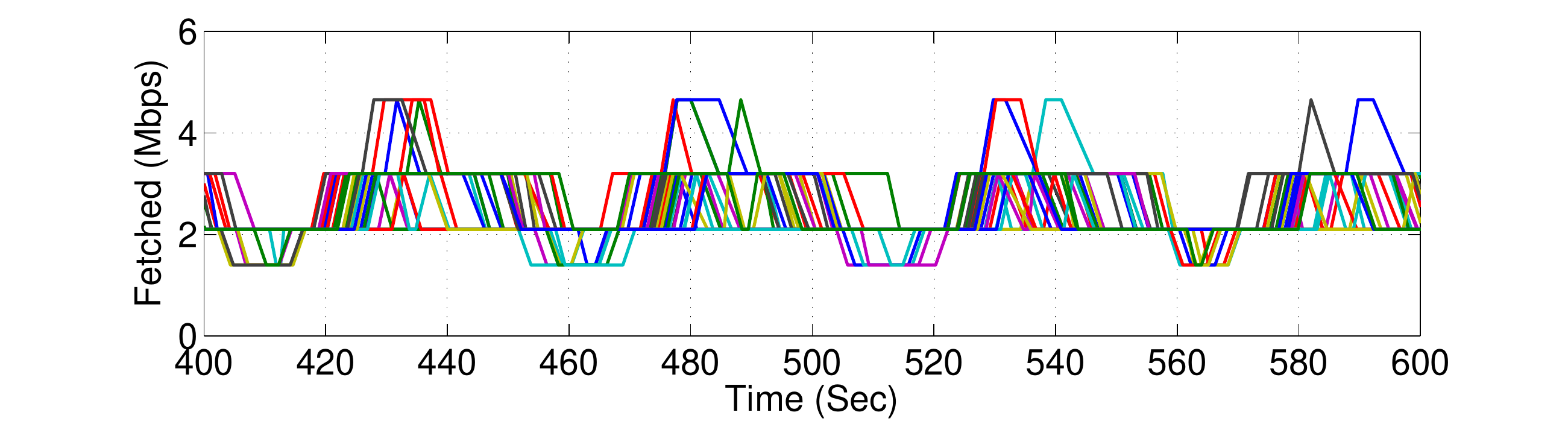}
\par\end{center}%
\end{minipage}
\par\end{centering}

\begin{centering}
\vspace{0in}

\par\end{centering}

\centering{}\caption{Oscillation of video bitrate when 36 Microsoft Smooth clients compete
at a 100-Mbps link. For more detailed experimental setup, refer to
$\S$\ref{sub:Experimental-Setup}. }

\label{Flo:36smooth} \vspace{-0.15in}
\end{figure}

\vspace{-0.07in}

\subsection{Emerging Issues}

A major trend in HAS use cases is its large-scale deployment in managed
networks by service providers, which typically leads to aggregating
multiple HAS streams in the aggregation/core network. For example,
an important scenario is that within a single household or a neighborhood,
several HAS flows belonging to one DOCSIS%
\footnote{Data over cable service interface specification.%
} bonding group compete for bandwidth. In the unmanaged wide-area Internet,
as HAS is growing to become a substantial fraction of the total traffic,
it will also become more and more common to have multiple HAS streams
compete for available bandwidth at any network bottlenecks. 

While a simple rate adaptation algorithm might work fairly well for
the case where a single HAS stream operates alone or shares bandwidth
with non-HAS traffic, recent studies \cite{Jiang:CoNext12,Akhshabi:NOSSDAV12}
have reported undesirable behaviors when multiple HAS streams compete
for bandwidth at a bottleneck link. For example, while studies have
suggested that significant video quality variation over time is undesirable
for a viewer's quality of experience \cite{Mok:WMUST2011}, in \cite{Jiang:CoNext12}
the authors reported unstable video bitrate selection and unfair bandwidth
sharing among three Microsoft Smooth clients sharing a 3-Mbps link.
In our own test bed experiments (see Figure \ref{Flo:36smooth}),
we observed significant and regular video bitrate oscillation when
multiple Microsoft Smooth clients share a bottleneck link. We also
found that oscillation behavior persists under a wide range of parameter
settings, including the number of players, link bandwidth, start time
of clients, heterogeneous RTTs, random early detection (RED) queueing
parameters, the use of weight fair queueing (WFQ), the presence of
moderate web-like cross traffic, etc. 

Our study shows that these HAS rate oscillation and instability behaviors
are not incidental -- they are simply \emph{symptoms} of a much more
fundamental limitation of the conventional HAS rate adaptation algorithms,
in which \emph{the TCP downloading throughput observed by a client
is directly equated to its fair share of the network bandwidth}. This
fundamental problem would also impact a HAS client's ability to avoid
buffer underrun when the bandwidth suddenly drops. In brief, the problem
derives from the discrete nature of HAS video bitrates. This makes
it impossible to always match the video bitrate to the network bandwidth,
resulting in undersubscription of the network bandwidth. Undersubscription
is typically coupled with clients' on-off downloading patterns. The
off-intervals then become a source of ambiguity for a client to correctly
perceive its fair share of the network bandwidth, thus preventing
the client from making accurate rate adaptation decisions%
\footnote{In \cite{Akhshabi:NOSSDAV12}, Akhshabi et al. have reached similar
conclusions. But they identify the off-intervals instead of the TCP
throughput-based measurement as the root cause. Their sequel work
\cite{Akhshabi:NOSSDAV13} attempts to tackle the problem from a very
different angle using traffic shaping.%
}.

\vspace{-0.07in}

\subsection{Overview of Solution}

To overcome this fundamental limitation, we envision a solution based
on a ``probe-and-adapt'' principle. In this approach, the TCP downloading
throughput is taken as an input \emph{only when} it is an accurate
indicator of the fair-share bandwidth. This usually happens when the
network is oversubscribed (or congested) and the off-intervals are
absent. In the presence of off-intervals, the algorithm constantly
\emph{probes}%
\footnote{By probing, we mean small trial increment of data rate, instead of
sending auxiliary piggybacking traffic.%
} the network bandwidth by incrementing its sending rate, and prepares
to back off once it experiences congestion. This new mechanism shares
the same spirit with TCP's congestion control, but it operates independently
at the application layer and at a per-segment rather than a per-RTT
time scale. We present PANDA (Probe AND Adapt) -- a client-side rate
adaptation algorithm -- as a specific implementation of this principle.

\begin{figure}
\begin{centering}
\begin{minipage}[t]{1\columnwidth}%
\begin{center}
\includegraphics[scale=0.35]{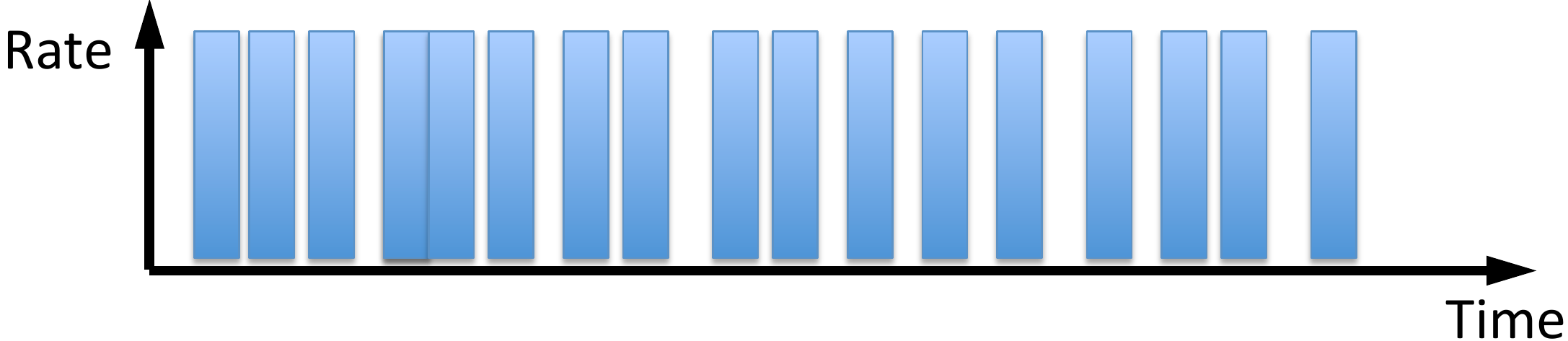} 
\par\end{center}

\begin{center}
\vspace{-0.2in}
(a) PANDA\vspace{-0.01in}

\par\end{center}%
\end{minipage}
\par\end{centering}

\begin{centering}
\vspace{0.05in}
\begin{minipage}[t]{1\columnwidth}%
\begin{center}
\includegraphics[scale=0.35]{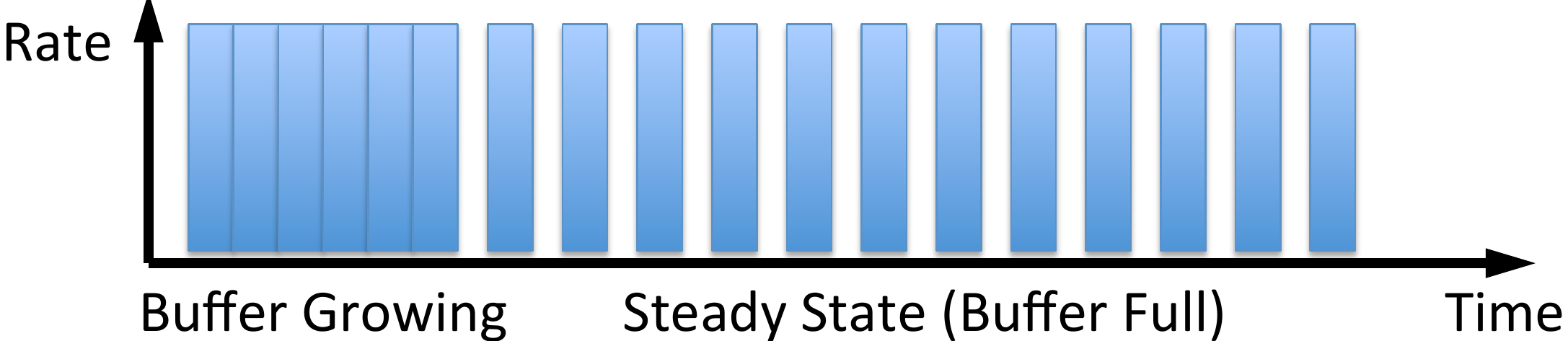} 
\par\end{center}

\begin{center}
\vspace{-0.05in}
(b) Conventional Bimodal 
\par\end{center}%
\end{minipage}
\par\end{centering}

\vspace{0.05in}
\caption{Illustration of PANDA's fine-granular request intervals vs. a conventional
algorithm's bimodal request intervals.}

\label{Flo:delay} \vspace{-0.05in}
\end{figure}

Probing constitutes fine-tuning the requested network data rate, with
continuous variation over a range. By nature, the available video
bitrates in HAS can only be discrete. A main challenge in our design
is to create a continuous decision space out of the discrete video
bitrate. To this end, we propose to \emph{fine-tune the intervals
between consecutive segment download requests} such that the \emph{average
data rate} sent over the network is a continuous variable (see Figure
\ref{Flo:delay} for an illustrative comparison with the conventional
scheme). Consequently, instead of directly tuning the video bitrate,
we probe the bandwidth based on the average data rate, which in turn
determines the selected video bitrate and the fine-granularity inter-request
time. 

There are various benefits associated with the probe-and-adapt approach.
First, it avoids the pitfall of inaccurate bandwidth estimation. Having
a robust bandwidth measurement to begin with gives the subsequent
operations improved discriminative power (for example, strong smoothing
of the bandwidth measurement is no longer required, leading to better
responsiveness). Second, with constant probing via incrementing the
rate, the network bandwidth can be more efficiently utilized. Third,
it ensures that the bandwidth sharing converges towards fair share
(i.e., the same or adjacent video bitrate) among competing clients.
Lastly, an innate feature of the probe-and-adapt approach is \emph{asymmetry
of rate shifting} -- PANDA is equipped with conservative rate level
upshift but more responsive downshift. Responsive downshift facilitates
fast recovery from sudden bandwidth drops, and thus can effectively
mitigate the danger of playout stalls caused by buffer underrun.

\begin{center}
\begin{table}[t]
\centering\scriptsize%
\begin{minipage}[t]{0.99\columnwidth}%
\begin{center}
{\small }%
\begin{tabular}{|c|l|}
\hline 
{\small Notation } & {\small Explanation}\tabularnewline
\hline 
{\small $w$ } & {\small Probing additive increase bitrate }\tabularnewline
{\small $\kappa$ } & {\small Probing convergence rate}\tabularnewline
{\small $\alpha$ } & {\small Smoothing convergence rate}\tabularnewline
$\beta$ & {\small Client buffer convergence rate}\tabularnewline
{\small $\Delta$} & {\small Quantization margin}\tabularnewline
$\epsilon$ & {\small Multiplicative safety margin}\tabularnewline
{\small $\tau$ } & {\small Video segment duration (in video time)}\tabularnewline
{\small $B$ } & {\small Client buffer duration (in video time)}\tabularnewline
{\small $B_{\min}$; $B_{\max}$ } & {\small Minimum/maximum client buffer duration}\tabularnewline
{\small $T$ } & {\small Actual inter-request time}\tabularnewline
{\small $\hat{T}$ } & {\small Target inter-request time}\tabularnewline
{\small $\tilde{T}$ } & {\small Segment download duration}\tabularnewline
{\small $x$ } & {\small Actual average data rate}\tabularnewline
{\small $\hat{x}$ } & {\small Target average data rate (or bandwidth share)}\tabularnewline
{\small $\hat{y}$ } & {\small Smoothed version of $\hat{x}$}\tabularnewline
{\small $\tilde{x}$ } & {\small $ $TCP throughput measured, $\tilde{x}:=\frac{r\cdot\tau}{\tilde{T}}$}\tabularnewline
{\small $\mathcal{R}$ } & {\small Set of video bitrates $\mathcal{R}:=\{R_{1},...,R_{L}\}$}\tabularnewline
{\small $r$ } & {\small Video bitrate available from $\mathcal{R}$}\tabularnewline
$S(\cdot)$ & {\small Rate smoothing function}\tabularnewline
$Q(\cdot)$ & {\small Video bitrate quantization function}\tabularnewline
\hline 
\end{tabular}
\par\end{center}%
\end{minipage}\caption{\label{tab:notation}Notations used in this paper}
\vspace{-0.15in}
\end{table}

\par\end{center}

\subsection{Paper Organization}

In the rest of the paper, after formalizing the problem ($\S$\ref{sec:Problem-Setup}),
we first introduce a method to characterize the conventional rate
adaptation algorithms ($\S$\ref{sec:Existing-Rate-Adaptation}),
based on which we analyze the root cause of its problems ($\S$\ref{sec:Analysis-of-Rule-of-Thumb}).
We then introduce our probe-and-adapt approach ($\S$\ref{sec:Proposed-Rate-Adaptation})
to directly address the root cause, and present the PANDA rate adaptation
algorithm as a concrete implementation of this idea. We provide comprehensive
performance evaluations ($\S$\ref{sec:Experimental-Results}). We
conclude the paper with final remarks and discussion of future work
($\S$\ref{sec:Conclusions}).

\section{Problem Model\label{sec:Problem-Setup}}

In this section, we formalize the problem by first describing a representative
HAS server-client interaction process. We then outline a four-step
model for an HAS rate adaptation algorithm. This will allow us to
compare the proposed PANDA algorithm with its conventional counterpart.
Table \ref{tab:notation} lists the main notations used in this paper.

\subsection{Process of HAS Server-Client Interaction}

Consider that a video stream is chopped into segments of $\tau$ seconds
each. Each segment has been pre-encoded at $L$ video bitrates, all
stored at a server. Denote by $\mathcal{R}:=\{R_{1},...,R_{L}\}$
the set of available video bitrates, with $0<R_{\ell}<R_{m}$ for
$\ell<m$.

\begin{figure}
\begin{centering}
\includegraphics[scale=0.38]{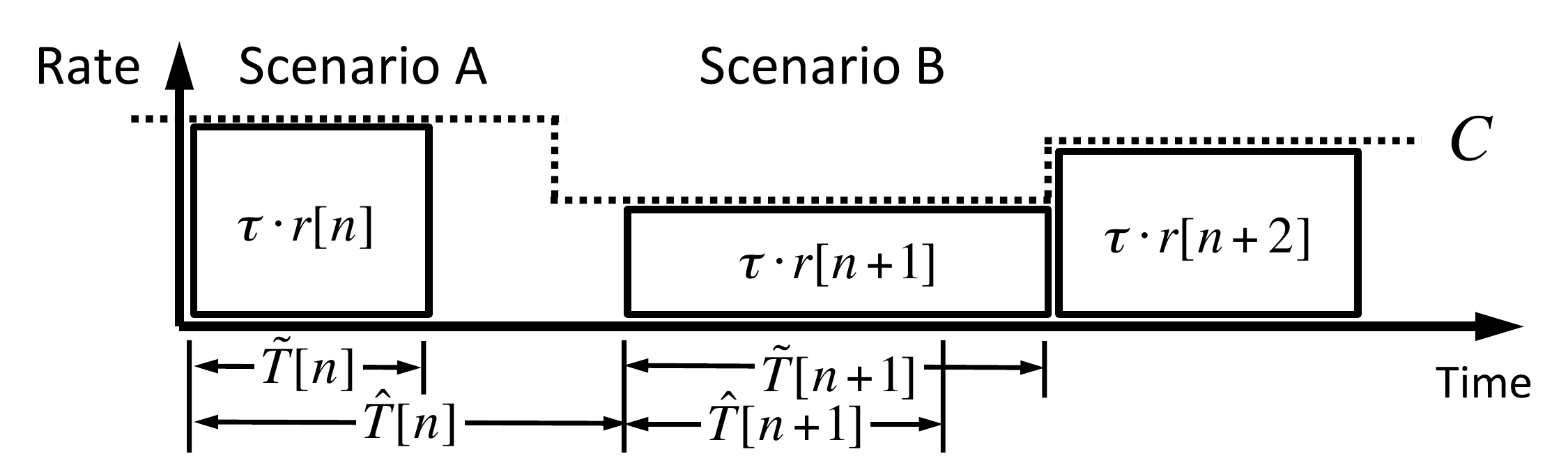} 
\par\end{centering}

\vspace{-0.1in}
\caption{The HAS segment downloading process.}

\label{Flo:delay2} \vspace{-0.15in}
\end{figure}

For each client, the streaming process is divided into sequential
segment downloading steps $n=1,2,...$. The process we consider here
generalizes the process used by conventional HAS clients by further
incorporating variable durations between consecutive segment requests.
Refer to Figure \ref{Flo:delay2}. At the beginning of each download
step $n$, a rate adaptation algorithm: 
\begin{itemize}
\item Selects the video bitrate of the next segment to be downloaded, $r[n]\in\mathcal{R}$;
\item Specifies how much time to give for the current download, until the
next download request (i.e., the inter-request time), $\hat{T}[n]$.
\end{itemize}
The client then initiates an HTTP GET request to the server for the
segment of sequence number $n$ and video bitrate $r[n]$, and the
downloading starts immediately. Let $\tilde{T}[n]$ be the\emph{ download
duration --} the time required to complete the download. Assuming
that no pipelining of downloading is involved, the next download step
starts after time 
\begin{equation}
T[n]=\max(\hat{T}[n],\tilde{T}[n]),\label{eq:tn}
\end{equation}
where $T[n]$ is the \emph{actual inter-request time}. That is, if
the download duration $\tilde{T}[n]$ is shorter than the target delay
$\hat{T}[n]$, the client waits time $\hat{T}[n]-\tilde{T}[n]$ (i.e.,
the off-interval) before starting the next downloading step (Scenario
A); otherwise, the client starts the next download step immediately
after the current download is completed (Scenario B).

Typically, a rate adaptation algorithm also measures its TCP throughput
$\tilde{x}$ during the segment downloading, via:

\[
\tilde{x}[n]:=\frac{r[n]\cdot\tau}{\tilde{T}[n]}.
\]

The downloaded segments are stored in the client buffer. After playout
starts, the buffer is consumed by the video player at a natural rate
of one video second per real second on average. Let $B[n]$ be the
buffer duration (measured in video time) at the end of step $n$.
Then the buffer dynamics can be characterized by:
\begin{equation}
B[n]=\max\left(0,B[n-1]+\tau-T[n]\right).\label{eq:buffer}
\end{equation}

\subsection{Four-Step Model\label{sub:Four-Step-Model}}

We present a four-step model for an HAS rate adaptation algorithm,
generic enough to encompass both the conventional algorithms (e.g.,
\cite{Liu:MMSys11,Tian:CoNext12,Zhou:VCIP12,Miller:PV12,Liu:SPIC12})
and the proposed PANDA algorithm. In this model, a rate adaptation
algorithm proceeds in the following four steps.
\begin{itemize}
\item \emph{Estimating}. The algorithm starts by estimating the network
bandwidth $\hat{x}[n]$ that can legitimately be used.
\item \emph{Smoothing}. $\hat{x}[n]$ is then noise-filtered to yield the
smoothed version $\hat{y}[n]$, with the aim of removing outliers.
\item \emph{Quantizing}. The continuous $\hat{y}[n]$ is then mapped to
the discrete video bitrate $r[n]\in\mathcal{R}$, possibly with the
help of side information such as client buffer size, etc.
\item \emph{Scheduling}. The algorithm selects the target interval until
the next download request, $\hat{T}[n]$$ $.
\end{itemize}

\section{Conventional Approach \label{sec:Existing-Rate-Adaptation}}

Using the four-step model above, in this section we introduce a scheme
to characterize a conventional rate adaptation algorithm, which will
serve as a benchmark.

To the best of our knowledge, almost all of today's commercial HAS
players%
\footnote{In this paper, the terms ``HAS player'' and ``HAS client'' are
used interchangeably.%
} implement the \emph{measuring} and \emph{scheduling} parts of the
rate adaptation algorithm in a similar way, though they may differ
in their implementation of the smoothing and quantizing parts of the
algorithm. Our claim is based on a number of experimental studies
of commercial HAS players \cite{akhashabi12SPIC,Huang:IMC12,Jiang:CoNext12}.
The scheme described in Algorithm \ref{alg:Baseline} characterizes
their essential ideas.

\begin{algorithm}
At the beginning of each downloading step $n$: 
\begin{enumerate}
\item Estimate the bandwidth share $\hat{x}[n]$ by equating it to the measured
TCP throughput: 
\begin{equation}
\hat{x}[n]=\tilde{x}[n-1].\label{eq:baseline2}
\end{equation}

\item Smooth out $\hat{x}[n]$ to produce filtered version $\hat{y}[n]$
by 
\begin{equation}
\hat{y}[n]=S(\{\hat{x}[m]:m\leq n\}).\label{eq:baseline1}
\end{equation}

\item Quantize $\hat{y}[n]$ to the discrete video bitrate $r[n]\in\mathcal{R}$
by 
\begin{equation}
r[n]=Q\left(\hat{y}[n];...\right).\label{eq:baseline4}
\end{equation}

\item Schedule the next download request depending on the buffer fullness:
\begin{equation}
\hat{T}[n]=\begin{cases}
0, & B[n-1]<B_{\max},\\
\tau, & \mathrm{otherwise.}
\end{cases}\label{eq:baseline3}
\end{equation}

\end{enumerate}
\caption{Conventional\label{alg:Baseline}}
\vspace{-0.1in}
\end{algorithm}

\begin{figure*}
\begin{centering}
\begin{minipage}[t]{1\columnwidth}%
\begin{flushleft}
\includegraphics[scale=0.27]{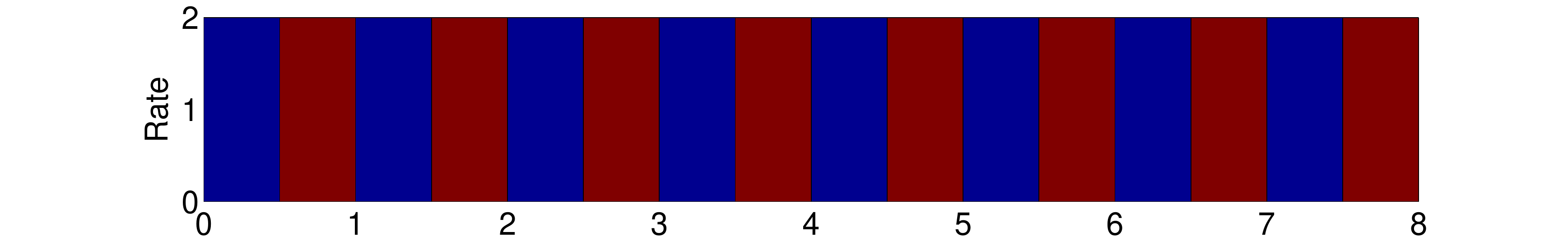} 
\par\end{flushleft}

\begin{flushleft}
\vspace{-0.1in}
\small\hspace{0.8in}(a) Perfectly Subscribed, Round-Robin\vspace{0.05in}

\par\end{flushleft}%
\end{minipage}%
\begin{minipage}[t]{1\columnwidth}%
\begin{flushleft}
\includegraphics[scale=0.27]{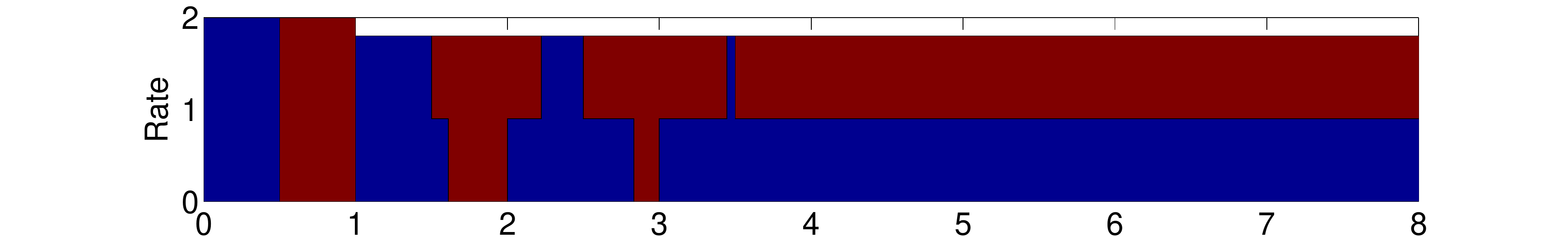} 
\par\end{flushleft}

\begin{flushleft}
\vspace{-0.1in}
\small\hspace{1.2in}(d) Oversubscribed
\par\end{flushleft}%
\end{minipage} %
\begin{minipage}[t]{1\columnwidth}%
\begin{flushleft}
\includegraphics[scale=0.27]{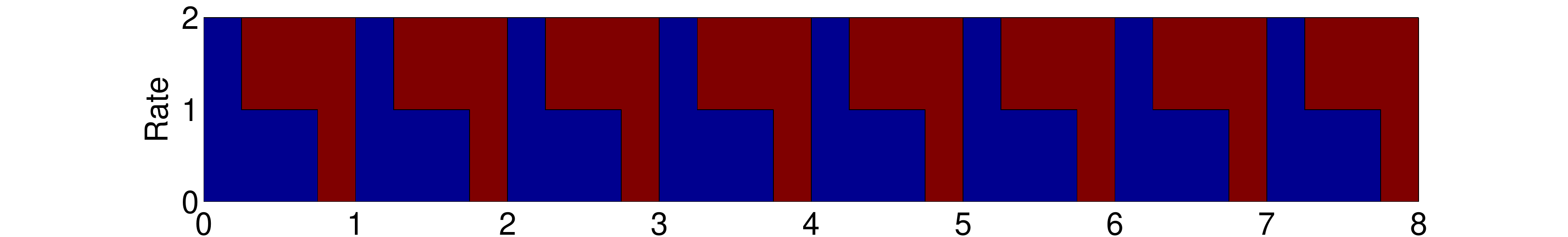} 
\par\end{flushleft}

\begin{flushleft}
\vspace{-0.1in}
\small\hspace{0.8in}(b) Perfectly Subscribed, Partially Overlapped
\vspace{0.05in}

\par\end{flushleft}%
\end{minipage}%
\begin{minipage}[t]{1\columnwidth}%
\begin{flushleft}
\includegraphics[scale=0.27]{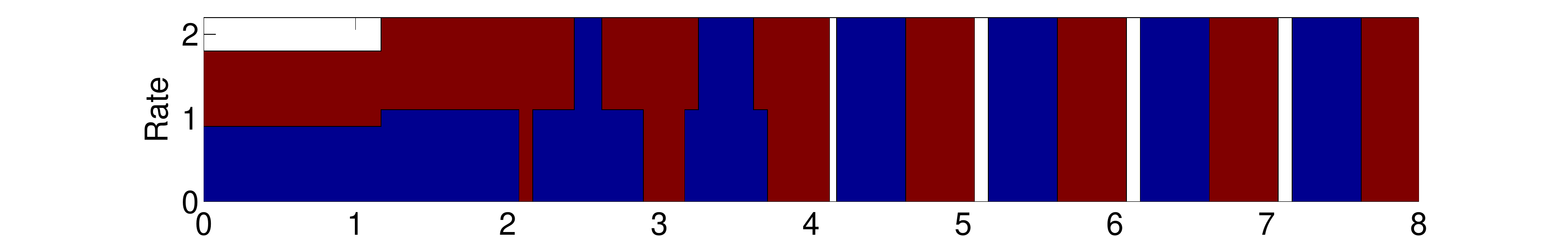} 
\par\end{flushleft}

\begin{flushleft}
\vspace{-0.1in}
\small\hspace{1.2in}(e) Undersubscribed \vspace{0.05in}

\par\end{flushleft}%
\end{minipage} %
\begin{minipage}[t]{1\columnwidth}%
\begin{flushleft}
\includegraphics[scale=0.27]{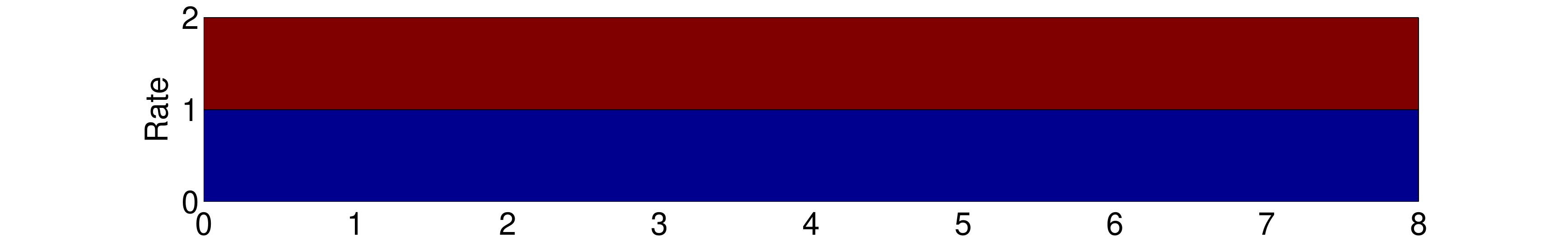} 
\par\end{flushleft}

\begin{flushleft}
\vspace{-0.1in}
\small\hspace{0.8in}(c) Perfectly Subscribed, Fully Overlapped \vspace{0.05in}

\par\end{flushleft}%
\end{minipage} %
\begin{minipage}[t]{1\columnwidth}%
\begin{flushleft}
\includegraphics[scale=0.27]{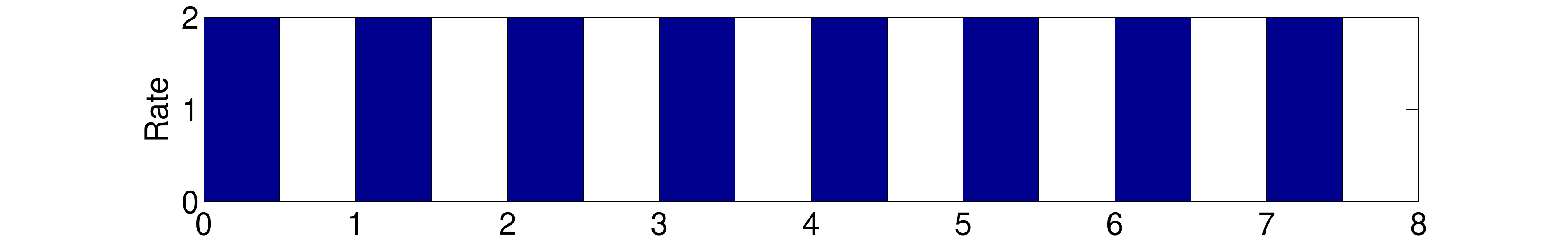} 
\par\end{flushleft}

\begin{flushleft}
\vspace{-0.1in}
\small\hspace{1.2in}(f) Single-Client
\par\end{flushleft}%
\end{minipage}
\par\end{centering}

\centering{}\caption{Illustration of various bandwidth sharing scenarios. In (a), (b) and
(c), the link is perfectly subscribed. In (d), the bandwidth sharing
starts with round-robin mode but then link becomes oversubscribed.
In (e), the bandwidth sharing starts with fully overlapped mode when
the link is oversubscribed. Starting from the second round, the link
becomes undersubscribed. In (f), a single client is downloading, and
the downloading on-off pattern exactly matches that of the blue segments
in (a).}

\label{Flo:bw_share} \vspace{-0.05in}
\end{figure*}

First, the algorithm equates the currently available bandwidth share
$\hat{x}[n]$ to the past TCP throughput $\tilde{x}[n-1]$ observed
during the on-interval $\tilde{T}[n-1]$. As the bandwidth is inferred
reactively based on the previous downloads, we refer to this as \emph{reactive
bandwidth estimation}.

The algorithm then obtains a filtered version $\hat{y}[n]$ using
a smoothing function $S(\cdot)$ that takes as input the measurement
history $\{\hat{x}[m]:m\leq n\}$, as described in (\ref{eq:baseline1}).
Various filtering methods are possible, such as sliding-window moving
average, exponential weighted moving average (EWMA) or harmonic mean
\cite{Jiang:CoNext12}.

The next step maps the continuous $ $$\hat{y}[n]$ to a discrete
video bitrate $r[n]\in\mathcal{R}$ using a quantization function
$Q(\cdot)$. In general, $Q(\cdot)$ can also incorporate side information,
including the past fetched bitrates $\{r[m]:m<n\}$ and the buffer
history $\{B[m]:m<n\}$. 

Lastly, the algorithm determines the target inter-request time $\hat{T}[n]$.
In (\ref{eq:baseline3}), $\hat{T}[n]$ is a mechanical function of
the buffer duration $B[n-1]$. If $B[n-1]$ is less than a pre-defined
maximum buffer $B_{\max}$, $\hat{T}[n]$ is set to $0$, and by (\ref{eq:tn}),
the next segment downloading starts right after the current download
is finished; otherwise, the inter-request time is set to the video
segment duration $\tau$, to stop the buffer from further growing.
This creates two distinct modes of segment downloading -- the \emph{buffer
growing} mode and the \emph{steady-state} mode, as shown in Figure
\ref{Flo:delay}(b). We refer to this as the \emph{bimodal download
scheduling}.

\section{Analysis of the Conventional Approach\label{sec:Analysis-of-Rule-of-Thumb}}

In this section, we take a deep dive into the conventional rate adaptation
algorithms and study their limitations.

\subsection{Bandwidth Cliff Effect\label{sub:Bandwidth-Overestimation}}

As we have seen in the previous section, conventional rate adaptation
algorithms use reactive bandwidth estimation (\ref{eq:baseline2})
that equates the estimated bandwidth share to the TCP throughput observed
during the on-intervals. In the presence of competing HAS clients,
however, the TCP throughput does not always faithfully represent the
fair-share bandwidth. In this section, we present an intuitive analysis
of this phenomenon, by extending the one first presented in \cite{Akhshabi:NOSSDAV12}.%
\footnote{A main difference of our analysis compared to \cite{Akhshabi:NOSSDAV12}
is that we rigorously prove the convergence properties presented in
the bandwidth cliff effect.%
} A rigorous analysis of this phenomenon is presented in Appendix \ref{sec:Bandwidth-Cliff-Effect:}.

First, we illustrate with simple examples. Figure \ref{Flo:bw_share}
(a) - (e) show the various scenarios of how a link can be shared by
two HAS clients in steady-state mode. We consider three different
scenarios: perfect link subscription, link oversubscription and link
undersubscription. We assume ideal TCP behavior, i.e., perfectly equal
sharing of the available bandwidth when the transfers overlap.

\emph{Perfect Subscription}: In perfect link subscription, the total
amount of traffic requested by the two clients perfectly fills the
link. (a), (b) and (c) illustrate three different modes of bandwidth
sharing, depending on the starting time of downloads relative to each
other. Essentially, under perfect subscription, there are unlimited
number of bandwidth sharing modes.

\emph{Oversubscription}: In (d), the two clients start with round-robin
mode and perfect subscription. Starting from the second round of downloading,
the bandwidth is reduced and the link becomes oversubscribed, i.e.,
each client requests segments larger than its current fair-share portion
of the bandwidth. This will result in unfinished downloads at the
end of each downloading round. Then, the unfinished segment will start
overlapping with segments of the next round. This repeats and the
downloading will become more and more overlapped, until all the clients
enter the fully overlapped mode.

\emph{Undersubscription}: In (e), initially the bandwidth sharing
is in fully overlapped mode, and the link is oversubscribed. Starting
from the second round, the bandwidth increases and the link becomes
undersubscribed. Then the clients start filling up each other's off-intervals,
until a transmission gap emerges. The bandwidth sharing will eventually
converge to a mode which is determined by the download start times. 

In any case, the measured TCP throughput faithfully represents the
fair-share bandwidth \emph{only when} the bandwidth sharing is in
the fully overlapped mode; in all other cases the TCP throughput overestimates
the fair-share bandwidth. Thus, most of the time, the bandwidth estimate
is accurate when the link is oversubscribed. Bandwidth overestimation
occurs when the link is undersubscribed or perfectly subscribed. In
general, when the number of competing clients is $n$, the bandwidth
overestimation ranges from one to $n$ times the fair-share bandwidth.

\begin{figure}
\begin{centering}
\includegraphics[scale=0.5]{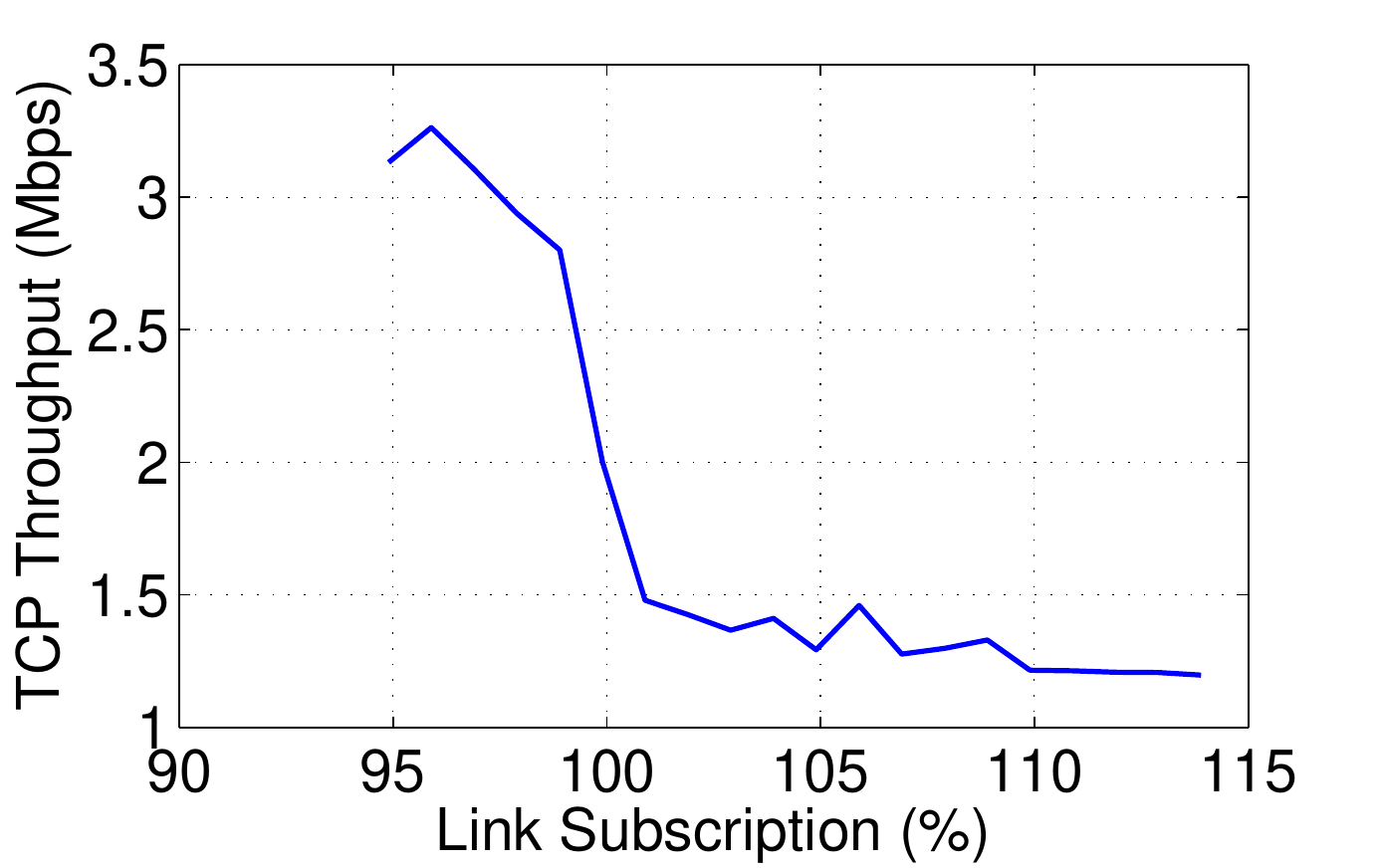}\vspace{-0.1in}

\par\end{centering}

\caption{Bandwidth cliff effect: measured TCP throughput vs. link subscription
rate for 100 thin clients sharing a 100-Mbps link. Each thin client
repeatedly downloads a segment every $\tau=2$ seconds.}

\label{Flo:bwcliff} \vspace{-0.05in}
\end{figure}

Although the preceding simple examples assume idealized TCP behavior
which abstracts away the complexity of TCP congestion control dynamics,
it is easy to verify that similar behavior occurs with real TCP connections.
To see this, we conducted a simple test bed experiment as follows.
We implemented a ``thin client'' to mimic an HAS client in the steady-state
mode. Each thin client repeatedly downloads a segment every 2 seconds.
We run 100 instances of the thin client sharing a bottleneck link
of 100 Mbps, each with a starting time randomly selected from a uniform
distribution between 0 and 2 seconds. Figure \ref{Flo:bwcliff} plots
the measured average TCP throughput as a function of the link subscription
rate. We observe that when the link subscription is below 100\%, the
measured throughput is about 3x the fair-share bandwidth of \textasciitilde{}1
Mbps. When the link subscription is above 100\%, the measured throughput
successfully predicts the fair-share bandwidth quite accurately. We
refer to this sudden transition from overestimation to fairly accurate
estimation of the bandwidth share at 100\% subscription as the \emph{bandwidth
cliff} \emph{effect}.

We summarize our findings as follows: 
\begin{itemize}
\item Link oversubscription converges to fully overlapped bandwidth sharing
and accurate bandwidth estimation. 
\item Link undersubscription converges to a bandwidth sharing pattern determined
by the download start times and bandwidth overestimation.
\item In perfect link subscription, there exist unlimited bandwidth sharing
modes, leading to bandwidth overestimation.
\end{itemize}

\subsection{Video Bitrate Oscillation}

With an understanding of the bandwidth cliff effect, we are now in
a good position to explain the bitrate oscillation observed in Figure
\ref{Flo:36smooth}. 

\begin{figure}
\begin{centering}
\vspace{0.1in}
\includegraphics[scale=0.33]{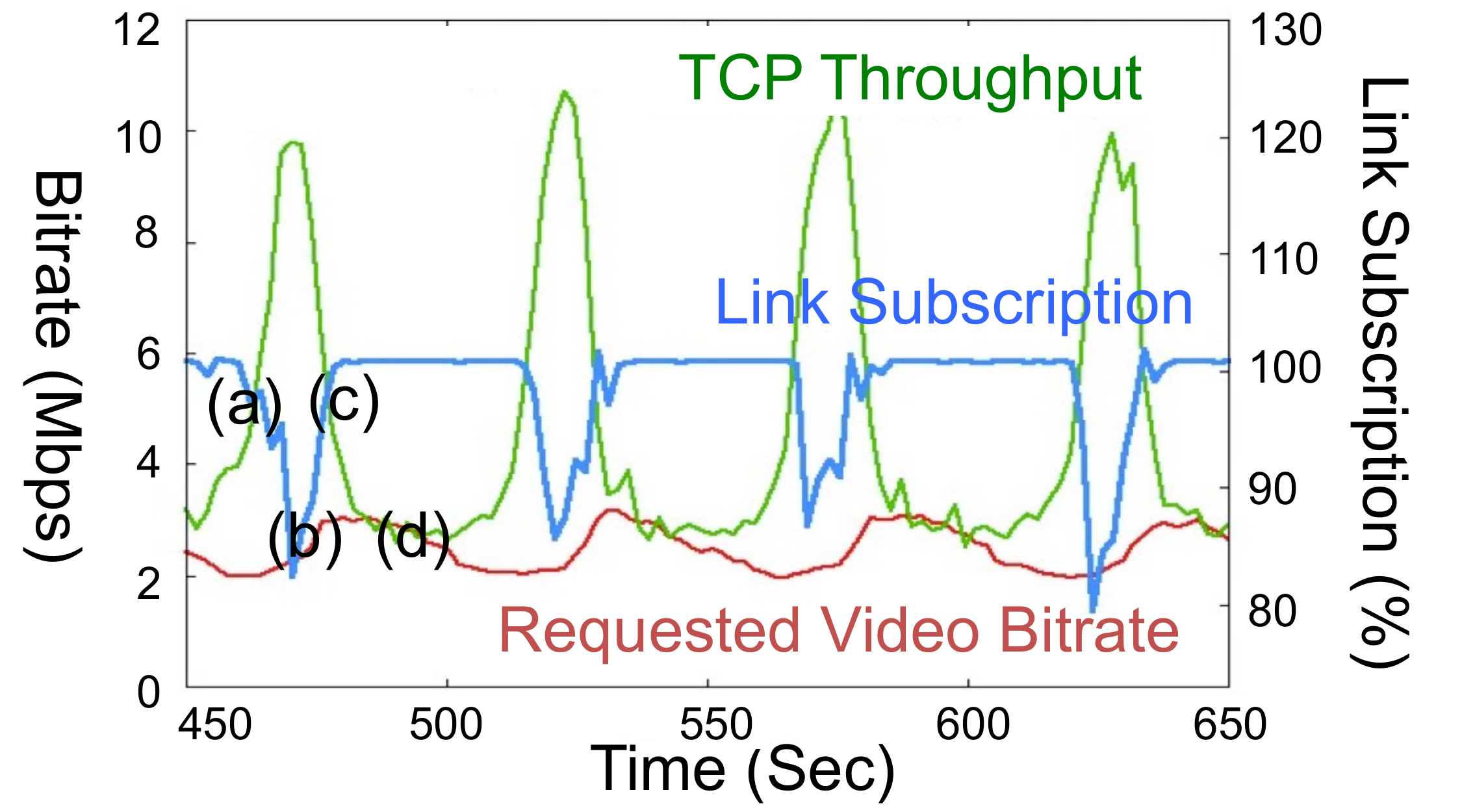}\vspace{-0.1in}

\par\end{centering}

\caption{Illustration of vicious cycle of video bitrate oscillation. This plot
is obtained with 36 Smooth clients sharing a 100-Mbps link. For experimental
setup, refer to $\S$\ref{sub:Experimental-Setup}. }

\label{Flo:oscillation} \vspace{-0.05in}
\end{figure}

Figure \ref{Flo:oscillation} illustrates this process. When the client
buffer reaches the maximum level \textbf{$B_{\max}$}, by (\ref{eq:baseline3}),
off-intervals start to emerge. The link becomes undersubscribed, leading
to bandwidth overestimation (a). This triggers the upshift of requested
video bitrate (b). As the available bandwidth cannot keep up with
the video bitrate, the buffer falls below \textbf{$B_{\max}$}. By
(\ref{eq:baseline3}), the client falls back to the buffer growing
mode and the off-intervals disappear, in which case the link again
becomes oversubscribed and the measured throughput starts to converge
to the fair-share bandwidth (c). Lastly, due to the quantization effect,
the requested video bitrate falls below the fair-share bandwidth (d),
and the client buffer starts growing again, completing one oscillation
cycle.

\subsection{Fundamental Limitation}

The bandwidth overestimation phenomenon reveals a more general and
fundamental limitation of the class of conventional reactive bandwidth
estimation approaches discussed so far. As video bitrates are chosen
solely based on measured TCP throughput from past segment downloads
during the on-intervals, such decisions completely ignore the network
conditions during the off-intervals. This leads to an \emph{ambiguity
of client knowledge} of available network bandwidth during the off-intervals,
which, in turn, hampers the adaptation process. 

To illustrate this point, consider two alternative scenarios as depicted
in Figures \ref{Flo:bw_share} (f) and (a). In (f), the client downloading
the blue (darker-shaded) video segments occupies the link alone; in
(a), it shares the same link with a competing client downloading the
green (lighter-shaded) video segments. Note that the on/off-intervals
for all the blue (darker-shaded) video segments follow exactly the
same pattern in both scenarios. Consequently, the client observes
exactly the same TCP throughput measurement over time. If the client
would obtain a complete picture of the network, it would know to upshift
its video bitrate in (f) but retain its current bitrate in (a). In
practice, however, an individual client cannot distinguish between
these two scenarios, hence, is bound to the same behavior in both. 

Note that as long as the off-intervals persist, such \emph{ambiguity
in client knowledge} is inherent to the bandwidth measurement step
in a network with competing streams. It cannot be resolved or remedied
by improved filtering, quantization, or scheduling steps performed
later in the client adaptation algorithm. Moreover, the bandwidth
cliff effect, as discussed in Section \ref{sub:Bandwidth-Overestimation},
suggests that the bandwidth overestimation problem does not improve
with more clients, and that it can introduce large errors even with
slight link undersubscription.

Instead, the client needs to take a more proactive approach in adapting
the video bitrate --- whenever it is known that the client knowledge
is impaired, it must \emph{avoid} using such knowledge in bandwidth
estimation. A way to distinguish the case when the knowledge is impaired
from when it is not, is to \emph{probe} the network subscription by
small increment of its data sending rate. We describe one algorithm
that follows such an alternative approach in the next section.

\section{Probe-and-Adapt Approach \label{sec:Proposed-Rate-Adaptation}}

In this section, we introduce our proposed probe-and-adapt approach
to directly address the root cause of the conventional algorithms'
problems. We begin the discussion by laying out the design goals that
a rate adaptation algorithm aims to achieve. We then describe the
PANDA algorithm as an embodiment of the probe-and-adapt approach,
and provide its functional verification using experimental traces.

\subsection{Design Goals\label{sec:Goals}}

Designing an HAS rate adaptation algorithm involves tradeoffs among
a number of competing goals. It is not legitimate to optimize one
goal (e.g., stability) without considering its tradeoff factors. From
an end-user's perspective, an HAS rate adaptation algorithm should
be designed to meet these criteria:
\begin{itemize}
\item \emph{Avoiding} \emph{buffer underrun}. Once the playout starts, buffer
underrun (i.e., complete depletion of buffer) leads to a playout stall.
Empirical study \cite{zhanghui2011} has shown that buffer underrun
may have the most severe impact on a user's viewing experience. To
avoid it, some minimal buffer level must be maintained at all times%
\footnote{Note that, however, the buffer level must also have an upper bound,
for a few different reasons. In live streaming, the end-to-end latency
from the real-time event to the event being displayed on user's screen
must be reasonably short. In video-on-demand, the maximum buffered
video must be limited to avoid wasted network usage in case of an
early termination of playback and to limit memory usage.%
}, and the adaptation algorithm must be highly responsive to network
bandwidth drops.
\item \emph{High quality smoothness}. In the simplest setting without considering
visual perceptual models, high video quality smoothness translates
into avoiding both frequent and significant video bitrate shifts among
available video bitrate levels \cite{Jiang:CoNext12,Mok:WMUST2011}. 
\item \emph{High average quality}. High average video quality dictates that
a client should fetch high-bitrate segments as much as possible. Given
a fixed network bandwidth, this translates into high network utilization.
\item \emph{Fairness}. In the simplest setting, fairness translates into
equal network bandwidth sharing among competing clients. 
\end{itemize}
Note that this list above is non-exhaustive. Other criteria, such
as low playout startup latency, are also important factors impacting
user's viewing experience.

\subsection{PANDA Algorithm}

In this section, we discuss the PANDA algorithm. Compared to the reactive
bandwidth estimation used by a conventional rate adaptation algorithm,
PANDA uses a more proactive probing mechanism. By probing, PANDA determines
a target average data rate $\hat{x}$. This average data rate is subsequently
used to determine the video bitrate $r$ to be fetched, and the interval
$\hat{T}$ until the next segment download request.

\begin{figure*}
\begin{centering}
\includegraphics[scale=0.38]{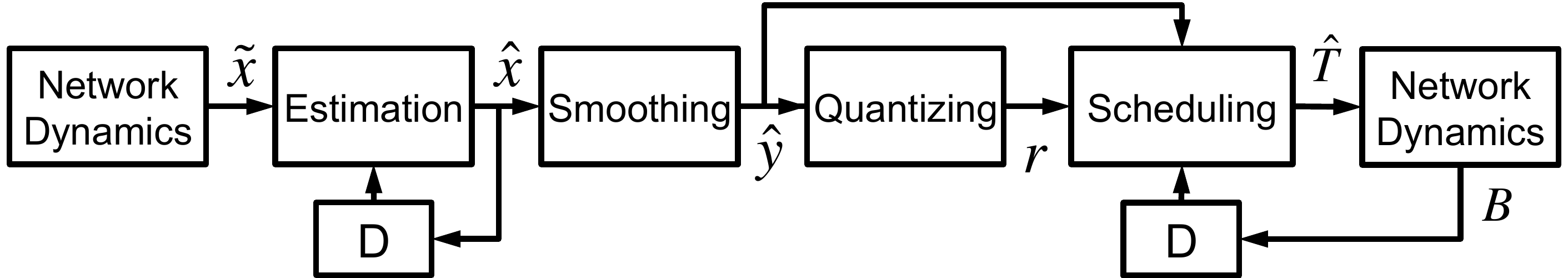}\vspace{-0.07in}

\par\end{centering}

\caption{Block diagram for PANDA (Algorithm \ref{alg:panda}). Module D represents
delay of one adaptation step.}

\label{Flo:panda_diagram} \vspace{-0.05in}
\end{figure*}

The PANDA algorithm is described in Algorithm \ref{alg:panda}, and
a block diagram interpretation of the algorithm is shown in Figure
\ref{Flo:panda_diagram}. Compared to the conventional algorithm in
Algorithm \ref{alg:Baseline}, we only make modifications in the \emph{estimating}
and \emph{scheduling} steps -- we replace (\ref{eq:baseline2}) with
(\ref{eq:baseline2b}) for estimating the bandwidth share, and (\ref{eq:baseline3})
with (\ref{eq:baseline3b}) for scheduling the next download request.
We now focus on elaborating each of these two modifications.

\begin{algorithm}
At the beginning of each downloading step $n$: 
\begin{enumerate}
\item Estimate the bandwidth share $\hat{x}[n]$ by
\begin{equation}
\frac{\hat{x}[n]-\hat{x}[n-1]}{T[n-1]}=\kappa\cdot(w-\max(0,\hat{x}[n-1]-\tilde{x}[n-1])),\label{eq:baseline2b}
\end{equation}

\item Smooth out $\hat{x}[n]$ to produce filtered version $\hat{y}[n]$
by 
\begin{equation}
\hat{y}[n]=S(\{\hat{x}[m]:m\leq n\}).\label{eq:baseline1b}
\end{equation}

\item Quantize $\hat{y}[n]$ to the discrete video bitrate $r[n]\in\mathcal{R}$
by 
\begin{equation}
r[n]=Q\left(\hat{y}[n];...\right).\label{eq:baseline4b}
\end{equation}

\item Schedule the next download request via 
\begin{equation}
\hat{T}[n]=\frac{r[n]\cdot\tau}{\hat{y}[n]}+\beta\cdot\left(B[n-1]-B_{\min}\right)\label{eq:baseline3b}
\end{equation}

\end{enumerate}
\caption{PANDA\label{alg:panda}}
\vspace{-0.1in}
\end{algorithm}

In the estimating step, (\ref{eq:baseline2b}) is designed to directly
address the root cause that leads to the video bitrate oscillation
phenomenon. Based on the insights obtained from $\S$\ref{sub:Bandwidth-Overestimation},
when the link becomes undersubscribed, the direct TCP throughput estimate
$\tilde{x}$ becomes inaccurate in predicting the fair-share bandwidth,
and thus should be avoided. Instead, the client continuously increments
the target average data rate $\hat{x}$ by $\kappa\cdot w$ per unit
time as a probe of the available capacity. Here $\kappa$ is the probing
convergence rate and $w$ is the additive increase rate. The algorithm
keeps on monitoring the TCP throughput $\tilde{x}$, and compares
it against the target average data rate $\hat{x}$. If $\tilde{x}>\hat{x}$,
$\tilde{x}$ would not be informative, since in this case the link
may still be undersubscribed and $\tilde{x}$ may overestimate the
fair-share bandwidth. Thus, its impact is suppressed by the $\max(0,\cdot)$
function. But if $\tilde{x}<\hat{x}$, then TCP throughput cannot
keep up with the target average data rate indicates that congestion
has occurred. This is when the target data rate $\hat{x}$ should
back off. The reduction imposed on $\hat{x}$ is made proportional
to $\hat{x}-\tilde{x}$. Intuitively, the lower the measured TCP throughput
$\tilde{x}$, the more reduction that needs to be imposed on $\hat{x}$.
This design makes our rate adaptation algorithm very agile to bandwidth
changes.

PANDA's probing mechanism shares similarities with TCP's congestion
control \cite{Jacobson1988}, and has an additive-increase-multiplicative-decrease
(AIMD) interpretation: $\kappa\cdot w$ is the additive increase term,
and $-\kappa\cdot\max(0,\hat{x}[n-1]-\tilde{x}[n-1])$ can be interpreted
as the multiplicative decrease term. The main difference is that in
TCP, congestion is indicated by packet losses (TCP Reno) or increased
round-trip time (delay-based TCP), whereas in (\ref{eq:baseline2b}),
congestion is indicated by the reduction of measured TCP throughput.
This AIMD property ensures that PANDA is able to efficiently utilize
the network bandwidth, and in the presence of multiple clients, the
bandwidth for each client eventually converges to fair-share status%
\footnote{Assuming the underlying TCP is fair (e.g., equal RTTs).%
}.

In the scheduling step, (\ref{eq:baseline3b}) aims to determine the
target inter-request time $\hat{T}[n]$. By right, $\hat{T}[n]$ should
be selected such that the smoothed target average data rate $\hat{y}[n]$
is equal to $\frac{r[n]\cdot\tau}{\hat{T}[n]}$. But additionally,
the selection of $\hat{T}[n]$ should also drive the buffer $B[n]$
towards a minimum reference level $B_{\min}>0$, so the second term
is added to the right hand side of (\ref{eq:baseline3b}), where $\beta>0$
controls the convergence rate.

One distinctive feature of the PANDA algorithm is its hybrid closed-loop/open-loop
design. Refer to Figure \ref{Flo:panda_diagram}. In this system,
(\ref{eq:baseline2b}) forms a closed loop by itself that determines
the target average data rate $\hat{x}$. (\ref{eq:baseline3b}) forms
a closed loop by itself that determines the target inter-request time
$\hat{T}$. Overall, the estimating, smoothing, quantizing and scheduling
steps together form an open loop. The main motivation behind this
design is to reduce the bitrate shifts associated with quantization.
Since quantization is excluded from the closed loop of $\hat{x}$,
it allows $\hat{x}$ to settle in a steady state. Since $r[n]$ is
a deterministic function of $\hat{x}[n]$, it can also settle in a
steady state.

In Appendix \ref{sec:Analysis-of-PANDA}, we present an equilibrium
and stability analysis of PANDA. We summarize the main results as
follows. Our equilibrium analysis shows that at steady state, the
system variables settle at
\begin{eqnarray}
\hat{x}_{o} & = & \tilde{x}_{o}+w\label{eq:ss_xhat}\\
 & = & \hat{y}_{o}\nonumber \\
r_{o} & = & Q(\hat{x}_{o};...)\nonumber \\
B_{o} & = & \left(1-\frac{r_{o}}{\hat{y}_{o}}\right)\cdot\frac{\tau}{\beta}+B_{\min},\label{eq:ss_B}
\end{eqnarray}
where the subscript $o$ denotes value of variables at equilibrium.
Our stability analysis shows that for the system to converge towards
the steady state, it is necessary to have:
\begin{eqnarray}
\text{\ensuremath{\kappa}} & < & \frac{2}{\tau}\label{eq:stability_k}\\
w & \leq & \Delta,\label{eq:stability_w}
\end{eqnarray}
where $\Delta$ is a parameter associated with the quantizer $Q(\cdot)$,
referred to as the \emph{quantization margin}, i.e., the selected
discrete rate $r$ must satisfy
\begin{equation}
r[n]\leq\hat{y}[n]-\Delta.\label{eq:rydelta}
\end{equation}

\subsection{Functional Verification}

We verify the behavior of PANDA using experimental traces. For detailed
experiment setup (including the selection of function $S(\cdot)$
and $Q(\cdot)$), refer to $\S$\ref{sub:Experimental-Setup}. 

\begin{figure}
\begin{centering}
\includegraphics[scale=0.36]{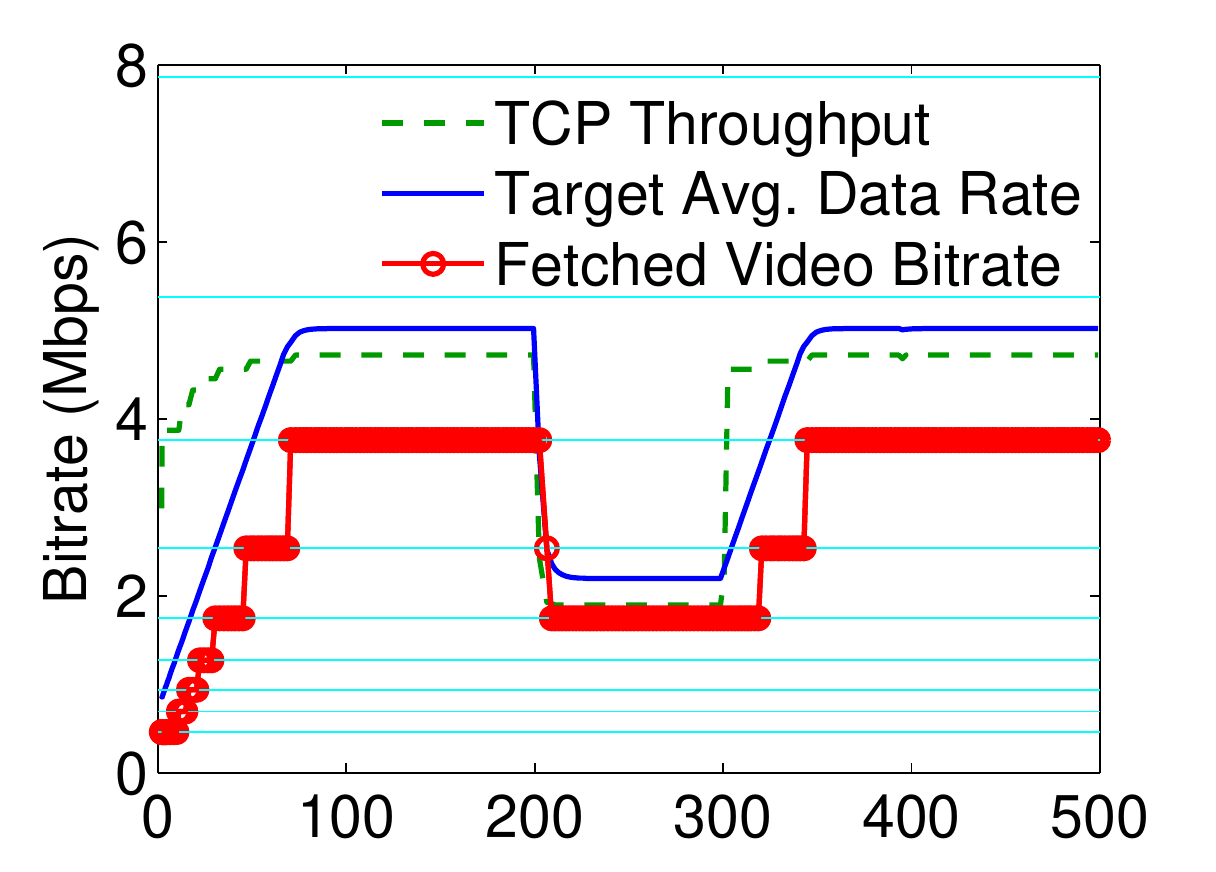}\includegraphics[scale=0.36]{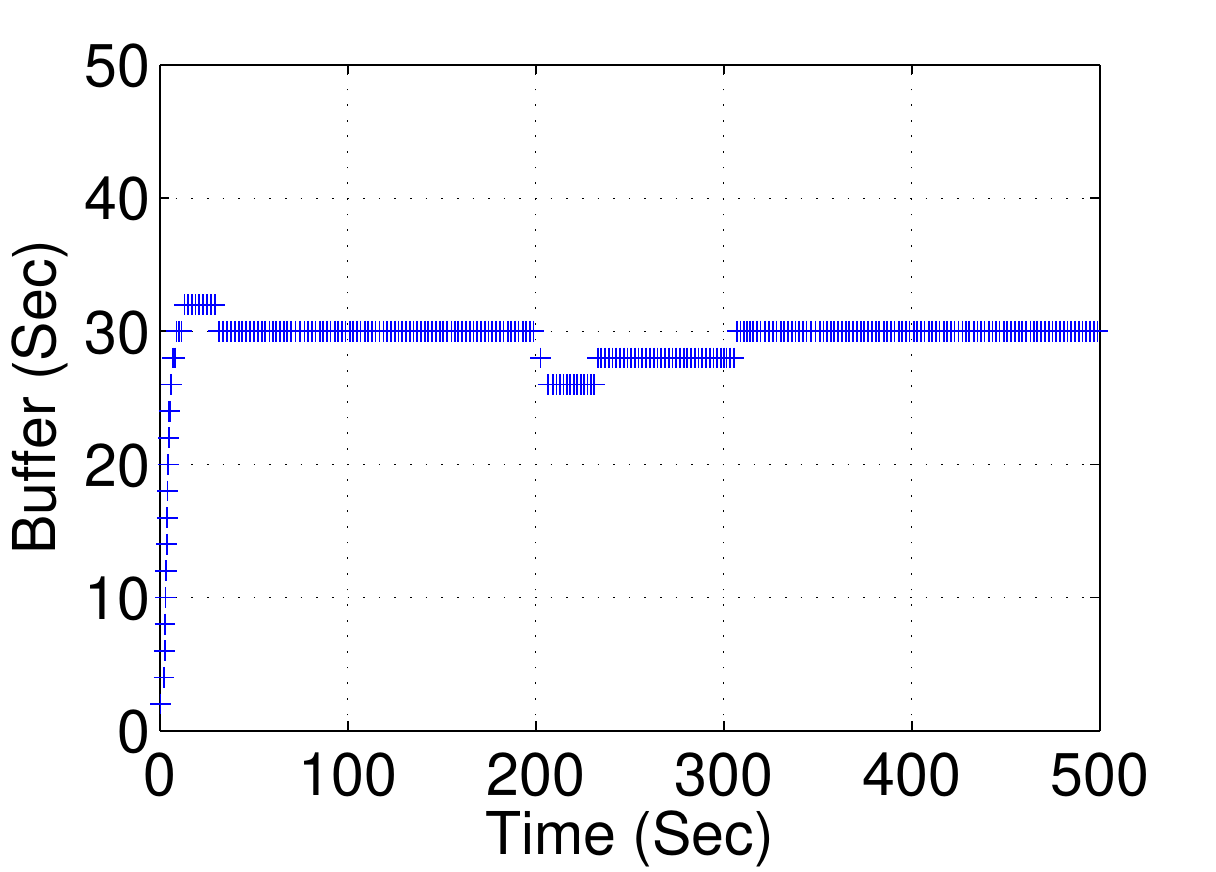}
\vspace{-0.23in}

\par\end{centering}

\caption{A PANDA client adapts its video bitrate under a bandwidth-varying
link. The bandwidth is initially at 5 Mbps, drops to 2 Mbps at 200
seconds and rises back to 5 Mbps at 300 seconds. }

\label{Flo:plainPANDA} \vspace{-0.05in}
\end{figure}

First, we evaluate how a single PANDA client adjusts its video bitrate
as the the available bandwidth varies over time. In Figure \ref{Flo:plainPANDA},
we plot the TCP throughput $\tilde{x}$, the target average data rate
$\hat{x}$, the fetched video bitrate $r$ and the client buffer $B$
for a duration of 500 seconds, where the bandwidth drops from 5 to
2 Mbps at 200 seconds, and rises back to 5 Mbps at 300 seconds. Initially,
the target average data rate $\hat{x}$ ramps up gradually over time;
the fetched video bitrate $r$ also ramps up correspondingly. After
the initial ramp-up stage, $\hat{x}$ settles in a steady state. It
can be observed that at steady state, the difference between $\hat{x}$
and $ $$\tilde{x}$ is about 0.3 Mbps, equal to $w$, which is consistent
with (\ref{eq:ss_xhat}). Similarly, the buffer $B$ also settles
in a steady state, and after plugging in all the parameters, one can
verify that the steady state of buffer (\ref{eq:ss_B}) also holds.
At 200 seconds, when the bandwidth suddenly drops, the fetched video
bitrate quickly drops to the desirable level. With this quick response,
the buffer hardly drops. This property makes PANDA favorable for live
streaming applications. When the bandwidth rises back to 5 Mbps at
300 seconds, the fetched video bitrate gradually ramps up to the original
level.

\begin{figure}
\begin{centering}
\includegraphics[scale=0.36]{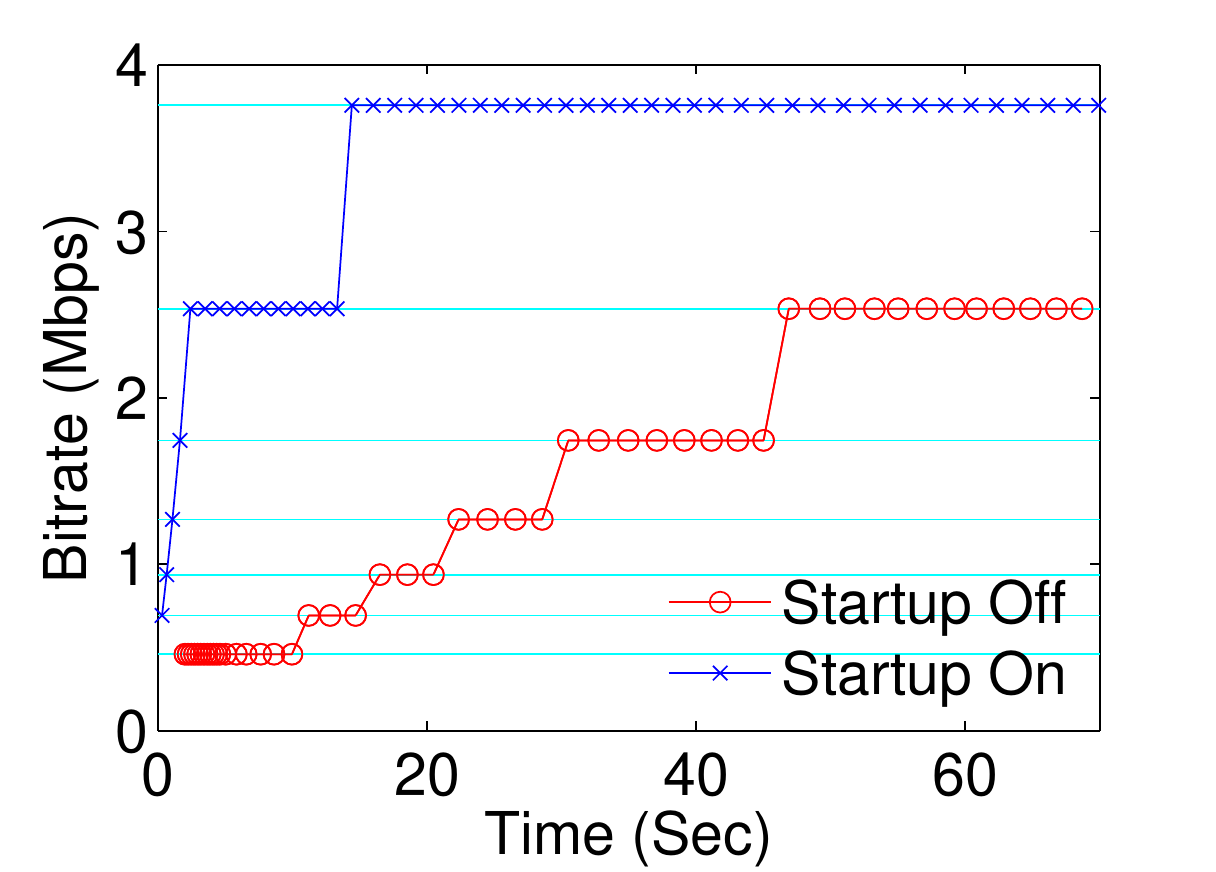}\includegraphics[scale=0.36]{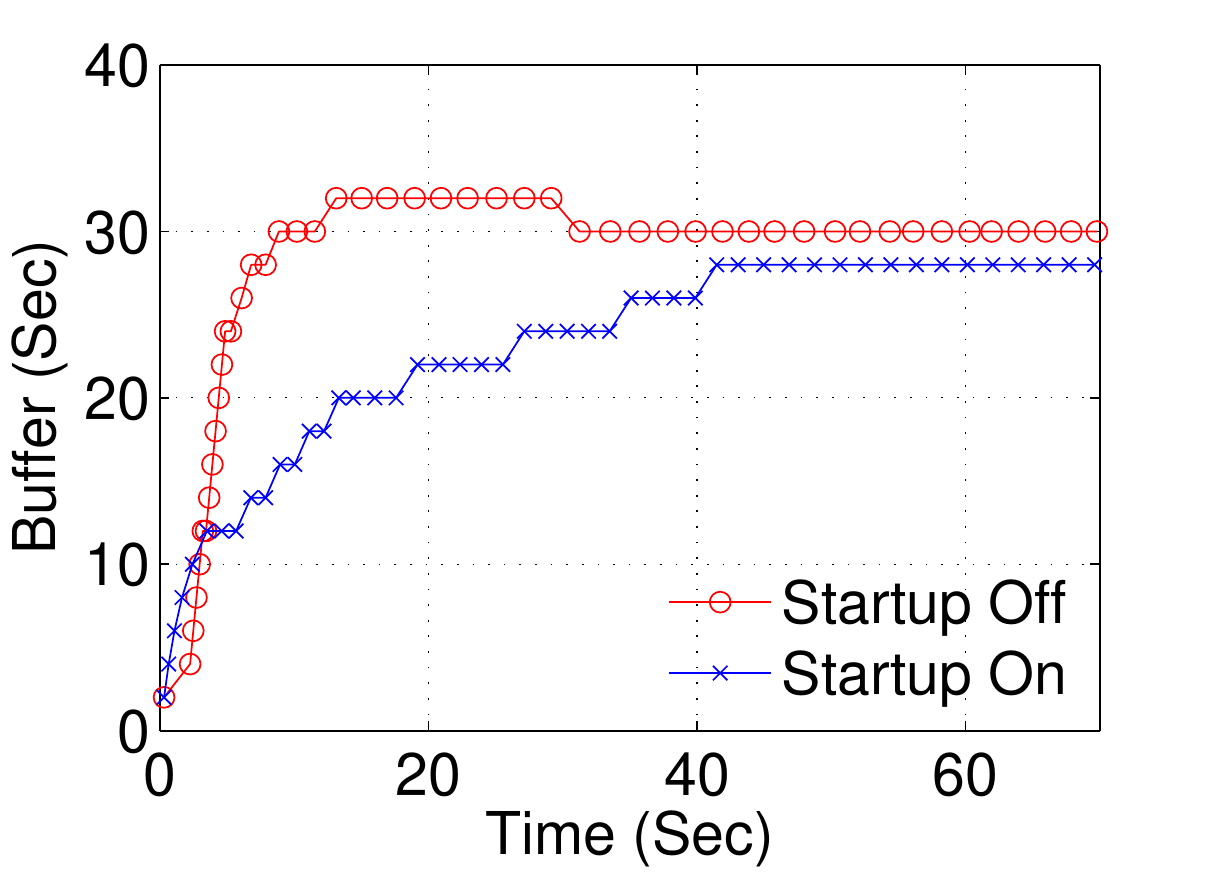}
\vspace{-0.23in}

\par\end{centering}

\caption{Comparison of the startup behavior of a PANDA player with and without
the startup logic. The bandwidth is 5 Mbps.}

\label{Flo:startup} \vspace{-0.05in}
\end{figure}

Note that, in practical implementation, we can further add a startup
logic to improve PANDA's ramp-up speed at the stream startup stage,
akin to the slow-start mode of TCP. The idea is simple: since it is
necessary to add off-intervals \emph{only when} the buffer duration
$B[n]$ exceeds the minimum reference level $B_{\min}$, we can use
the conventional algorithm at startup or after playout stall, until
$B[n]\geq B_{\min}$; after that, we switch to the main Algorithm
\ref{alg:panda}. Without the presence of the off-intervals, the conventional
algorithm works fast enough and reasonably well. Figure \ref{Flo:startup}
shows the startup behavior of a PANDA player with 5 Mbps link bandwidth,
with and without the startup logic. As can be seen, the startup logic
allows the video bitrate to ramp up efficiently, albeit at the expense
of somewhat dampened buffer growth.

\begin{figure}
\begin{centering}
\hspace{-0.3in} %
\begin{minipage}[t]{1\columnwidth}%
\begin{center}
\hspace{0.2in}\footnotesize Fetched Bitrate Aggregated over 36 Streams\vspace{-0.23in}

\par\end{center}

\begin{center}
\includegraphics[scale=0.35]{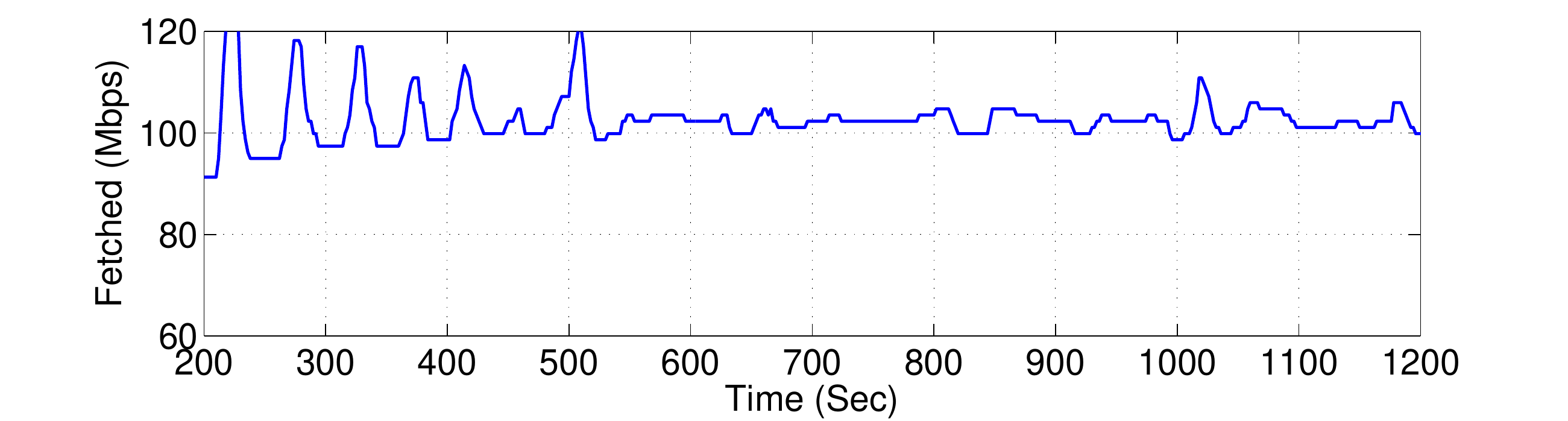}
\par\end{center}%
\end{minipage}
\par\end{centering}

\begin{centering}
\vspace{0.05in}
\hspace{-0.3in} %
\begin{minipage}[t]{1\columnwidth}%
\begin{center}
\hspace{0.2in}\footnotesize Fetched Bitrate of Individual Streams
(Zoom In)\vspace{-0.23in}

\par\end{center}

\begin{center}
\includegraphics[scale=0.35]{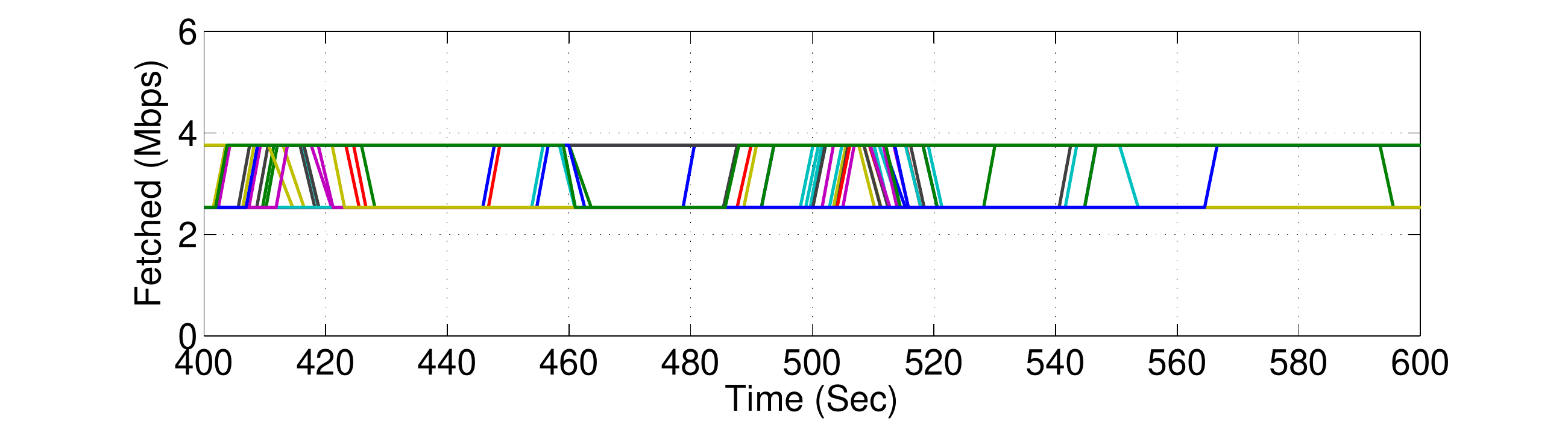}
\par\end{center}%
\end{minipage}
\par\end{centering}

\begin{centering}
\vspace{0in}

\par\end{centering}

\caption{36 PANDA clients compete at a 100-Mbps link in steady state. }

\label{Flo:36panda} \vspace{-0.05in}
\end{figure}

The more intriguing question is whether PANDA could effectively stop
the bitrate oscillation observed in the Smooth players. We conduct
an experiment with the same setup as the experiment shown in Figure
\ref{Flo:36smooth}, except that the PANDA player and the Smooth player
use slightly different video bitrate levels (due to different packaging
methods). The resulting fetched bitrates in aggregate and for each
client are shown in Figure \ref{Flo:36panda}. From the plot of the
aggregate fetched bitrate, except for the initial fluctuation, the
aggregate bitrate closely tracks the available bandwidth of 100 Mbps.
Zooming in to the individual streams' fetched bitrates, the fetched
bitrates are confined within two adjacent bitrate levels and the number
of shifts is much smaller than the Smooth client's case. This affirms
that PANDA is able to achieve better stability than the Smooth's rate
adaptation algorithm. In $\S$\ref{sec:Experimental-Results}, we
perform a comprehensive performance evaluation on each adaptation
algorithm.

\begin{figure}
\begin{centering}
\hspace{-0.3in} %
\begin{minipage}[t]{1\columnwidth}%
\begin{center}
\hspace{0.2in}\footnotesize Measured TCP Throughput $\tilde{x}$\vspace{-0.23in}

\par\end{center}

\begin{center}
\includegraphics[scale=0.35]{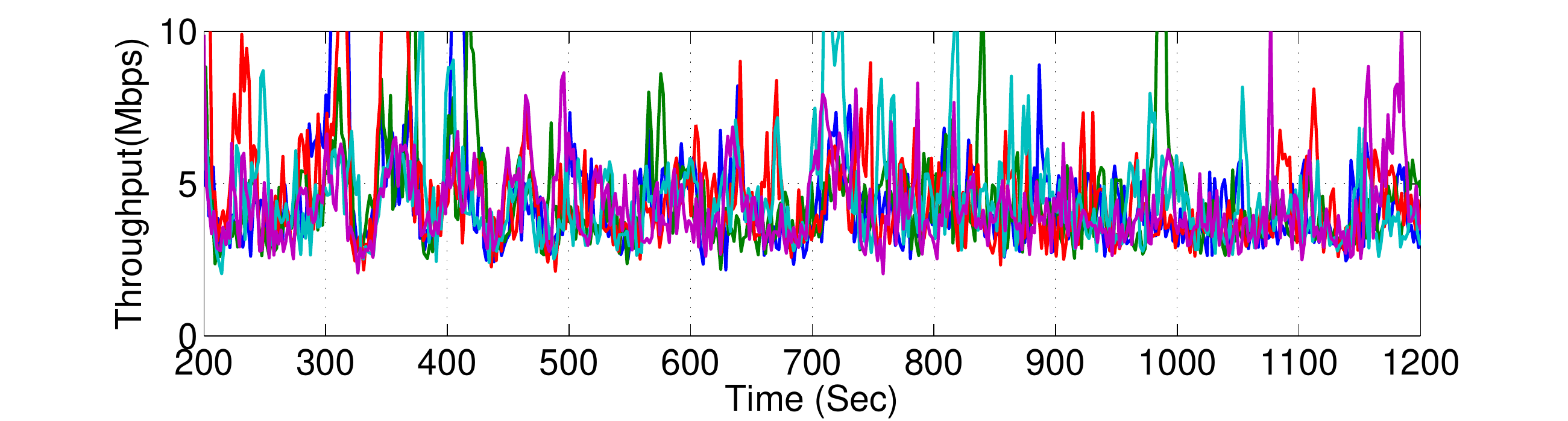}
\par\end{center}%
\end{minipage}
\par\end{centering}

\begin{centering}
\vspace{0.05in}
\hspace{-0.3in} %
\begin{minipage}[t]{1\columnwidth}%
\begin{center}
\hspace{0.2in}\footnotesize Target Average Data Rate $\hat{x}$\vspace{-0.23in}

\par\end{center}

\begin{center}
\includegraphics[scale=0.35]{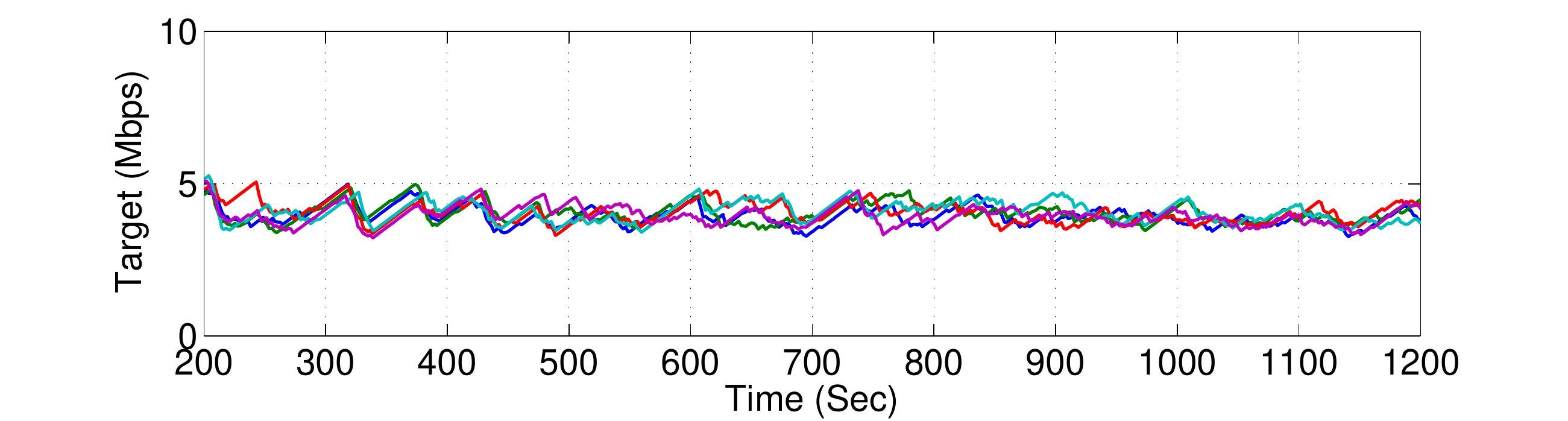}
\par\end{center}%
\end{minipage}
\par\end{centering}

\centering{}\caption{The traces of the TCP throughput and the target average data rate
of 36 PANDA clients compete at a 100-Mbps link in steady state. The
traces of the first five clients are plotted.}

\label{Flo:36panda_tput} \vspace{-0.05in}
\end{figure}

To help the reader develop a better intuition on why PANDA performs
better than a conventional algorithm, in Figure \ref{Flo:36panda_tput}
we plot the trace of the measured TCP throughput and the target average
data rate for the same experiment as in Figure \ref{Flo:36panda}.
Note that the fair-share bandwidth for each client is about 2.8 Mbps.
From the plot, the TCP throughput not only grossly overestimates the
fair-share bandwidth, it also has a large variation. If used directly,
this degree of noisiness gives the subsequent operations a very hard
job to extract useful information. For example, one may apply strong
filtering to smooth out the bandwidth measurement, but this would
seriously affect the responsiveness of the client. When the network
bandwidth suddenly drops, the client would not be able to respond
quickly enough to reduce its video bitrate, leading to catastrophic
buffer underrun. Moreover, the bias is both large and difficult to
predict, making any correction to the mean problematic. In comparison,
although also biased, the target average data rate estimated by the
probing mechanism is much less noisy than the TCP throughput. One
can easily correct the bias (via (\ref{eq:rydelta}) and quantization)
and select the right video bitrate without sacrificing responsiveness.

\begin{figure}
\begin{centering}
\begin{minipage}[t]{0.49\columnwidth}%
\begin{center}
\includegraphics[scale=0.36]{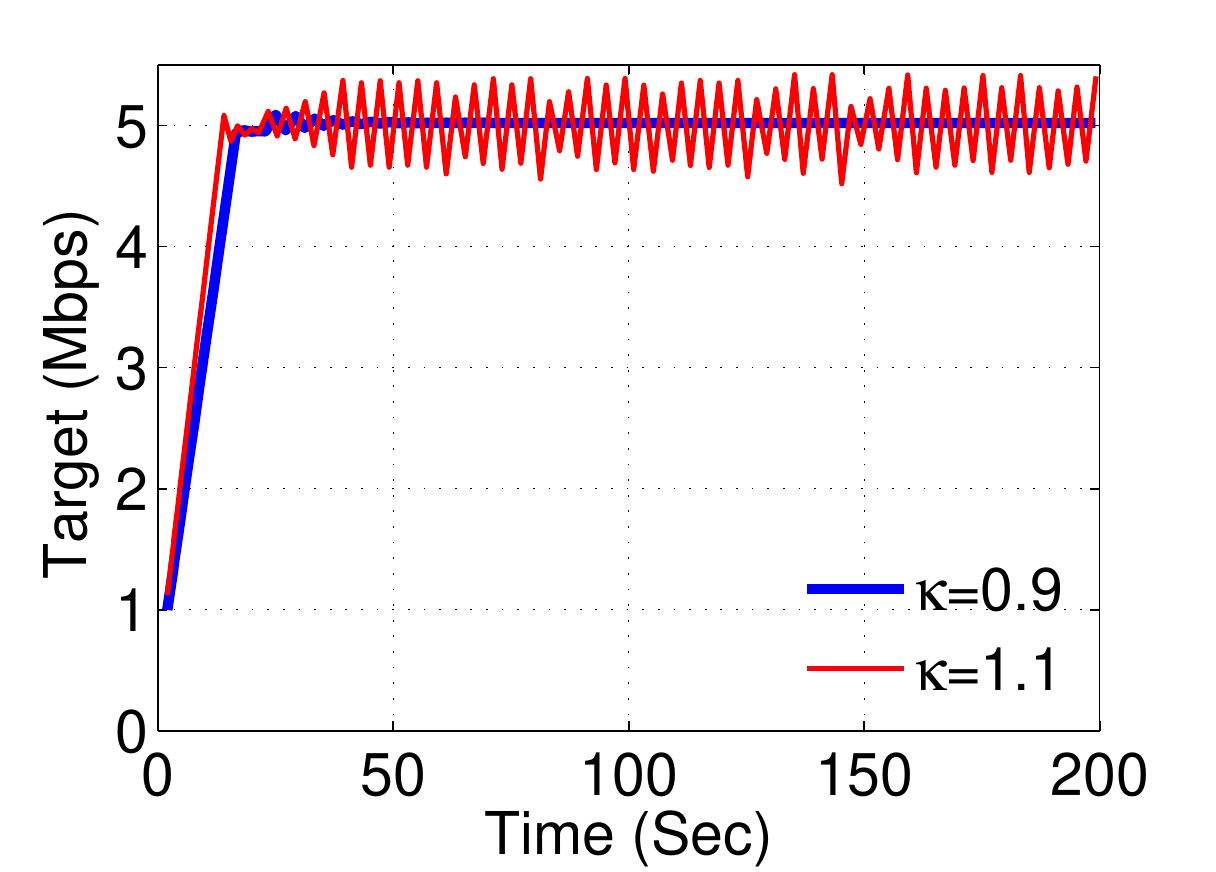}
\par\end{center}

\begin{center}
\vspace{-0.1in}
(a) $\kappa$
\par\end{center}

\begin{center}
\vspace{-0.15in}

\par\end{center}%
\end{minipage} %
\begin{minipage}[t]{0.49\columnwidth}%
\begin{center}
\includegraphics[scale=0.36]{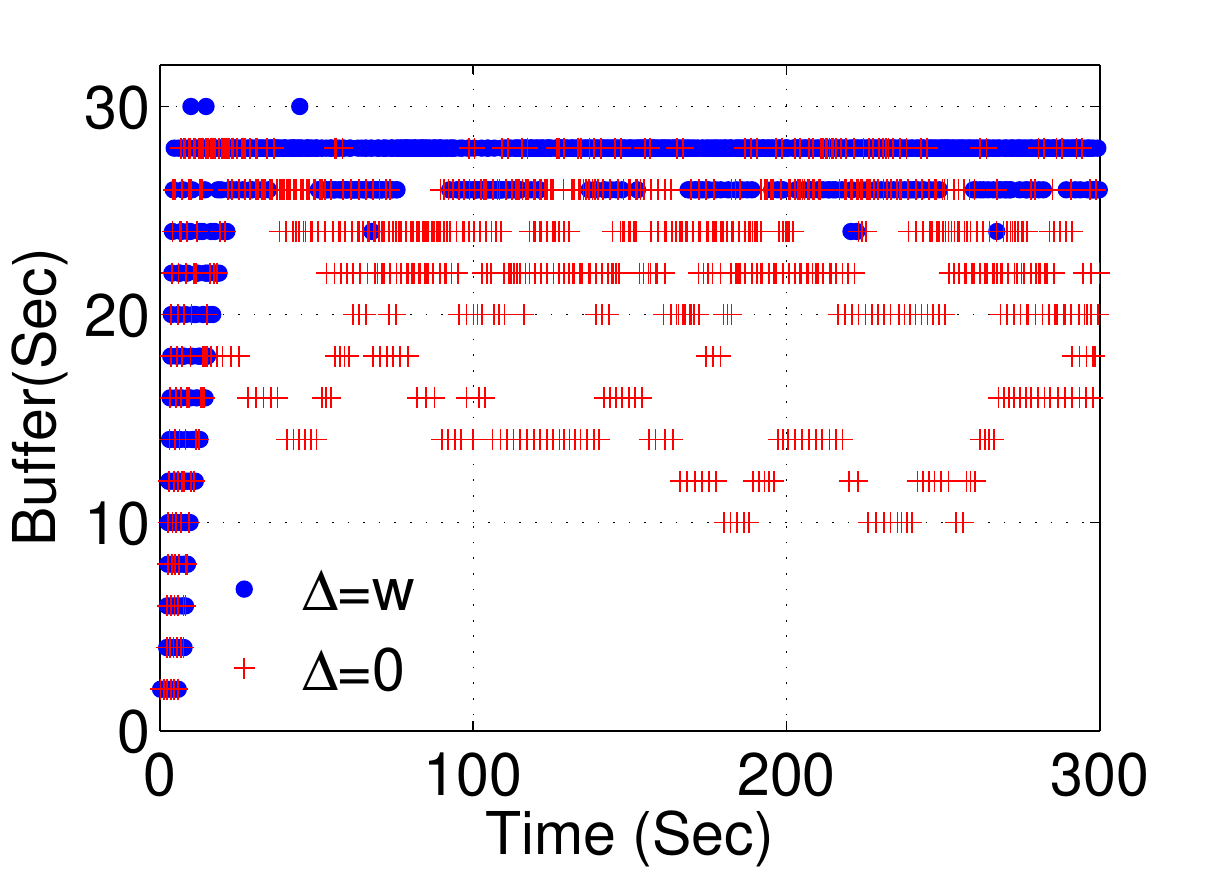}
\par\end{center}

\begin{center}
\vspace{-0.1in}
(b) $\Delta$
\par\end{center}

\begin{center}
\vspace{-0.15in}

\par\end{center}%
\end{minipage}
\par\end{centering}

\vspace{-0.07in}
\caption{Experimental verification of PANDA's stability criteria. In (a), one
PANDA client streams over a link of 5 Mbps. In (b), 10 PANDA clients
compete over a 10 Mbps link.}

\label{Flo:panda_stability} \vspace{-0.05in}
\end{figure}

In Figure \ref{Flo:panda_stability}, we verify the stability criteria
(\ref{eq:stability_k}) and (\ref{eq:stability_w}) of PANDA. With
$\tau=2$, the system is stable if $\kappa<1$. This is demonstrated
by Figure \ref{Flo:panda_stability} (a), where we show the traces
of the target average rate $\hat{x}$ for two $\kappa$ values 0.9
and 1.1. In Figure \ref{Flo:panda_stability} (b), we show that when
$\Delta=0$, the buffer cannot converge towards the reference level
of 30 seconds.

\section{Performance Evaluation\label{sec:Experimental-Results}}

In this section, we conduct a set of test bed experiments to evaluate
the performance of PANDA against other rate adaptation algorithms.

\subsection{Evaluation Metrics}

In $\S$\ref{sec:Goals}, we discussed four criteria that are most
important for a user's watching experience -- i) ability to avoid
buffer underruns, ii) quality smoothness, iii) average quality, and
iv) fairness. In this paper, we use \emph{buffer undershoot} as the
metric for Criterion i), described as follows.
\begin{itemize}
\item \emph{Buffer undershoot}: We measure how much the buffer goes down
after a bandwidth drop as a indicator of an algorithm's ability to
avoid buffer underruns. The less the buffer undershoot, the less likely
the buffer will underrun. Let $B_{o}$ be a reference buffer level
(30 seconds for all players in this paper), and $B_{i,t}$ the buffer
level for player $i$ at time $t$. The buffer undershoot for player
$i$ at time $t$ is defined as $\frac{\max(0,B_{o}-B_{i,t})}{B_{o}}$.
The buffer undershoot for player $i$ within a time interval $\mathcal{T}$
(right after a bandwidth drop), is defined as the 90th-percentile
value of the distribution of buffer undershoot samples collected during
$\mathcal{T}$. 
\end{itemize}
We inherit the metrics defined in \cite{Jiang:CoNext12} -- \emph{instability},
\emph{inefficiency} and \emph{unfairness}, as the metrics for Criteria
ii), iii) and iv), respectively. We only make a slight modification
to the definition of inefficiency. Let $r_{i,t}$ be the video bitrate
fetched by player $i$ at time $t$.
\begin{itemize}
\item \emph{Instability}: The instability for player $i$ at time $t$ is
$\frac{\sum_{d=0}^{k-1}|r_{i,t-d}-r_{i,t-d-1}|\cdot w(d)}{\sum_{d=0}^{k-1}r_{i,t-d}\cdot w(d)}$,
where $w(d)=k-d$ is a weight function that puts more weight on more
recent samples. $k$ is selected as 20 seconds.
\item \emph{Inefficiency}: Let $C$ be the available bandwidth. \cite{Jiang:CoNext12}
defines inefficiency as $\frac{\left|\sum_{i}r_{i,t}-C\right|}{C}$
for player $i$ at time $t$. But sometimes the sum of fetched bitrate
$\sum_{i}r_{i,t}$ can be greater than $C$. To avoid unnecessary
penalty in this case, we revise the inefficiency metric to $\frac{\max\left(0,C-\sum_{i}r_{i,t}\right)}{C}$
for player $i$ at time $t$.
\item \emph{Unfairness}: Let $JainFair_{t}$ be the Jain fairness index
calculated on the rates $r_{i,t}$ at time $t$ over all players.
The unfairness at $t$ is defined as $\sqrt{1-JainFair_{t}}$.
\end{itemize}

\subsection{Experimental Setup\label{sub:Experimental-Setup}}

\emph{HAS Player Configuration}: The benchmark players that we use
to compare against PANDA are:
\begin{itemize}
\item Microsoft Smooth player \cite{MSS}, a commercially available proprietary
player. The Smooth players are of version 1.0.837.34 using Silverlight
runtime version 4.0.50826. To our best knowledge, the Smooth player
as well as the Apple HLS and the Adobe HDS players all use the same
TCP throughput measurement mechanism, so we picked the Smooth player
as a representative.
\item FESTIVE player, which we implemented based on the details specified
in \cite{Jiang:CoNext12}. The FESTIVE algorithm is the first known
client-side rate adaptation algorithm designed to specifically address
the multi-client scenario.
\item A player implementing the conventional algorithm (Algorithm \ref{alg:Baseline}),
which differs from PANDA only in the estimating and scheduling steps.
\end{itemize}
For fairness, we ensure that PANDA and the conventional player use
the same smoothing and quantizing functions. For smoothing, we implemented
a \emph{EWMA} \emph{smoother} of the form: $\frac{\hat{y}[n]-\hat{y}[n-1]}{T[n-1]}=-\alpha\cdot\left(\hat{y}[n-1]-\hat{x}[n]\right)$,
where $\alpha>0$ is the convergence rate of $\hat{y}[n]$ towards
$\hat{x}[n]$. For quantization, we implemented a \emph{dead-zone
quantizer} $r[n]=Q(\hat{y}[n],r[n-1])$, defined as follows: Let the
upshift threshold be defined as $r_{up}:=\begin{array}{ccc}
\max_{r\in\mathcal{R}}r\end{array}$ subject to $r\leq\hat{y}[n]-\Delta_{up}$, and the downshift threshold
as $r_{down}:=\begin{array}{ccc}
\max_{r\in\mathcal{R}}r\end{array}$ subject to $r\leq\hat{y}[n]-\Delta_{down}$, where $\Delta_{up}$
and $\Delta_{down}$ are the upshift and downshift safety margin respectively,
with $\Delta_{up}\geq\Delta_{down}\geq0$. The dead-zone quantizer
updates $r[n]$ as 
\begin{equation}
r[n]=\begin{cases}
r_{up}, & r[n-1]<r_{up},\\
r[n-1], & r_{up}\leq r[n-1]\leq r_{down},\\
r_{down}, & \mathrm{otherwise}.
\end{cases}\label{eq:deadzone}
\end{equation}
The ``dead zone'' $[r_{up},r_{down}]$ created by setting $\Delta_{up}>\Delta_{down}$
mitigates frequent bitrate hopping between two adjacent levels, thus
stabilizing the video quality (i.e. hysteresis control). For the conventional
player, set $ $$\Delta_{up}=\epsilon\cdot\hat{y}$ and $\Delta_{down}=0$,
where $0\leq\epsilon<1$ is the multiplicative safety margin. For
PANDA, due to (\ref{eq:stability_w}) and (\ref{eq:rydelta}), set
$\Delta_{up}=w+\epsilon\cdot\hat{y}$ and $\Delta_{down}=w$ %
\footnote{Note that this will not give PANDA any unfair advantage.%
}.

Table \ref{tab:parameters} lists the default parameters used by each
player, as well as their varying values. For fairness, all players
attempt to maintain a steady-state buffer of 30 seconds. For PANDA,
$B_{\min}$ is selected to be 26 seconds such that the resulting steady-state
buffer is 30 seconds (by (\ref{eq:ss_B})).

\emph{Server Configuration}: The HTTP server runs Apache on Red Hat
6.2 (kernel version 2.6.32-220). The Smooth player interacts with
an Microsoft IIS server by default, but we also perform experiments
of Smooth player interacting with an Apache server on Ubuntu 10.04
(kernel version 2.6.32.21).

\begin{figure}
\begin{centering}
\includegraphics[scale=0.6]{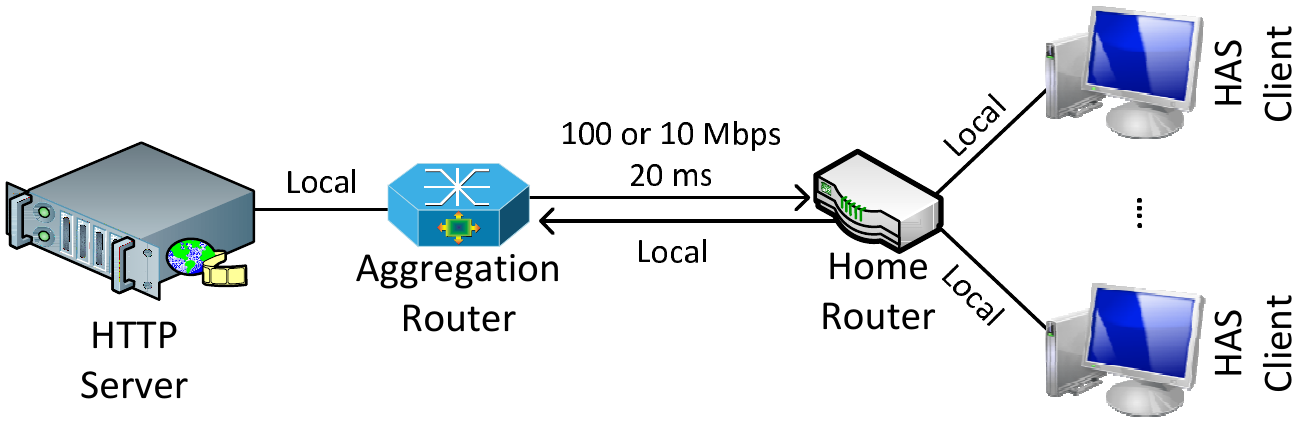} \vspace{-0.05in}

\par\end{centering}

\centering{}\caption{The network topology configured in the test bed. Local indicates that
the bitrate is effectively unbounded and the link delay is 0 ms.}

\label{Flo:testbed_network} \vspace{-0.1in}
\end{figure}

\emph{Network} \emph{Configuration}: As service provider deployment
over a managed network is our primary case of interest, our experiments
are configured to highly match the imporant scenario where a number
of HAS flows compete for bandwidth within a DOCSIS bonding group.
The test bed is configured as in Figure \ref{Flo:testbed_network}.
The queueing policy used at the aggregation router-home router bottleneck
link is the following. For a link bandwidth of 10 Mbps or below, we
use random early detection (RED) with $(min\_thr,max\_thr,p)=(30,90,0.25)$;
if the link bandwidth is 100 Mbps, we use RED with $(min\_thr,max\_thr,p)=(300,900,1)$.
The video content is chopped into segments of $\tau=2$ seconds, pre-encoded
with $L=10$ bitrates: 459, 693, 937, 1270, 1745, 2536, 3758, 5379,
7861 and 11321 Kbps. For the Smooth player, the data rates after packaging
are slightly different.

\begin{center}
\begin{table}[t]
\scriptsize\vspace{0.10in}%
\begin{minipage}[t]{0.99\columnwidth}%
\begin{center}
{\small }%
\begin{tabular}{|l|l|l|c|}
\hline 
Algorithm & Parameter & Default & Values\tabularnewline
\hline 
PANDA & $\kappa$ & 0.14 & 0.04,0.07,0.14,0.28,0.42,0.56\tabularnewline
 & $w$ & 0.3 & \tabularnewline
 & $\alpha$ & 0.2 & 0.05,0.1,0.2,0.3,0.4,0.5\tabularnewline
 & $\beta$ & 0.2 & \tabularnewline
 & $\epsilon$ & 0.15 & 0.5,0.4,0.3,0.2,0.1,0\tabularnewline
 & $B_{\min}$ & 26 & \tabularnewline
\hline 
Conventional & $\alpha$ & 0.2 & 0.01,0.04,0.07,0.1,0.15,0.2\tabularnewline
 & $\epsilon$ & 0.15 & \tabularnewline
 & $B_{\max}$ & 30 & \tabularnewline
\hline 
FESTIVE & Window & 20 & 20,15,10,6,3,1\tabularnewline
 & $targetbuf$ & 30 & \tabularnewline
\hline 
\end{tabular}
\par\end{center}%
\end{minipage}\caption{\label{tab:parameters}Parameters used in experiments}
\vspace{-0.05in}
\end{table}

\par\end{center}

\begin{figure*}
\begin{centering}
\begin{minipage}[t]{0.66\columnwidth}%
\begin{center}
\includegraphics[scale=0.35]{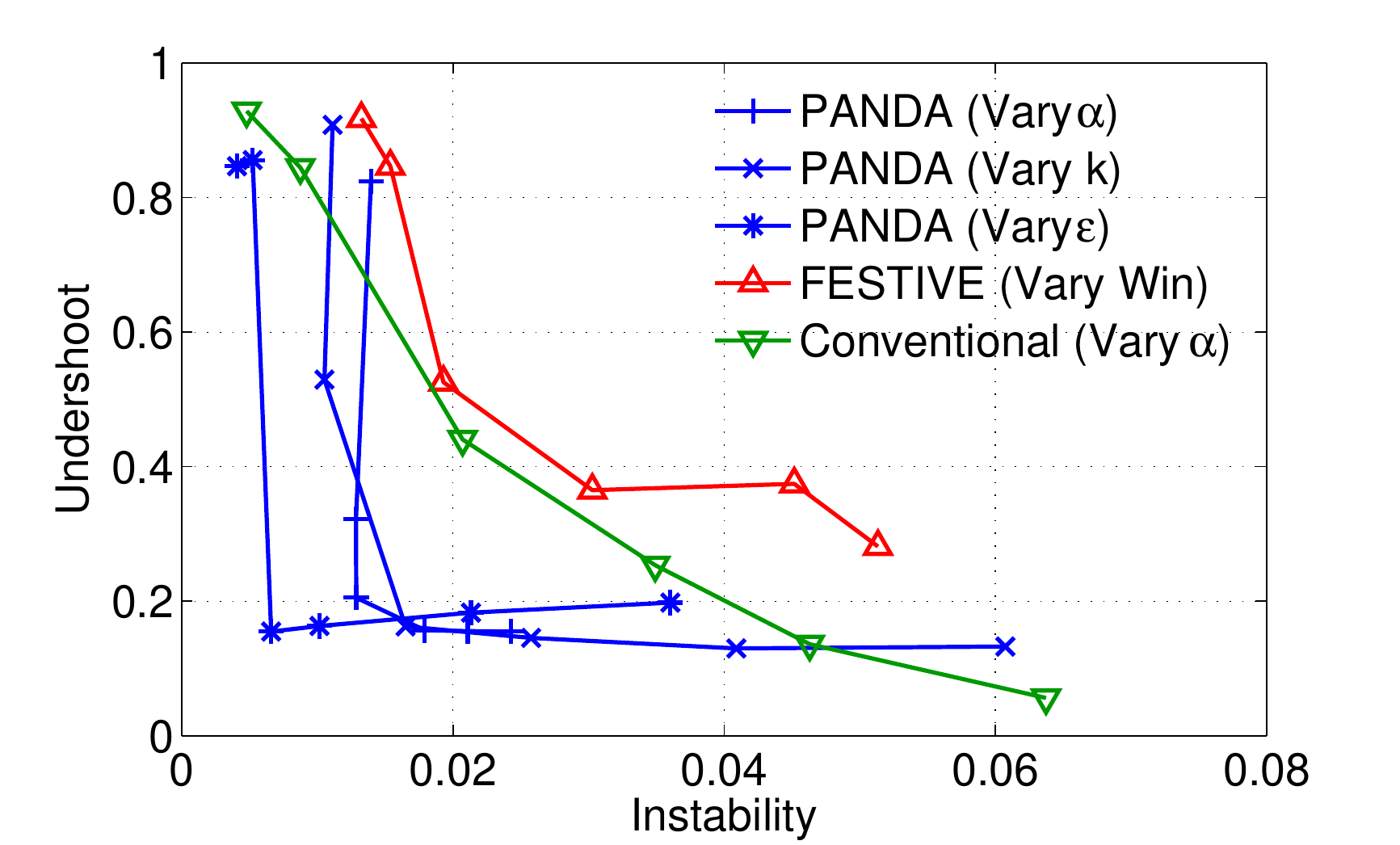}
\par\end{center}

\begin{center}
\vspace{-0.1in}
(a)
\par\end{center}%
\end{minipage} %
\begin{minipage}[t]{0.66\columnwidth}%
\begin{center}
\includegraphics[scale=0.35]{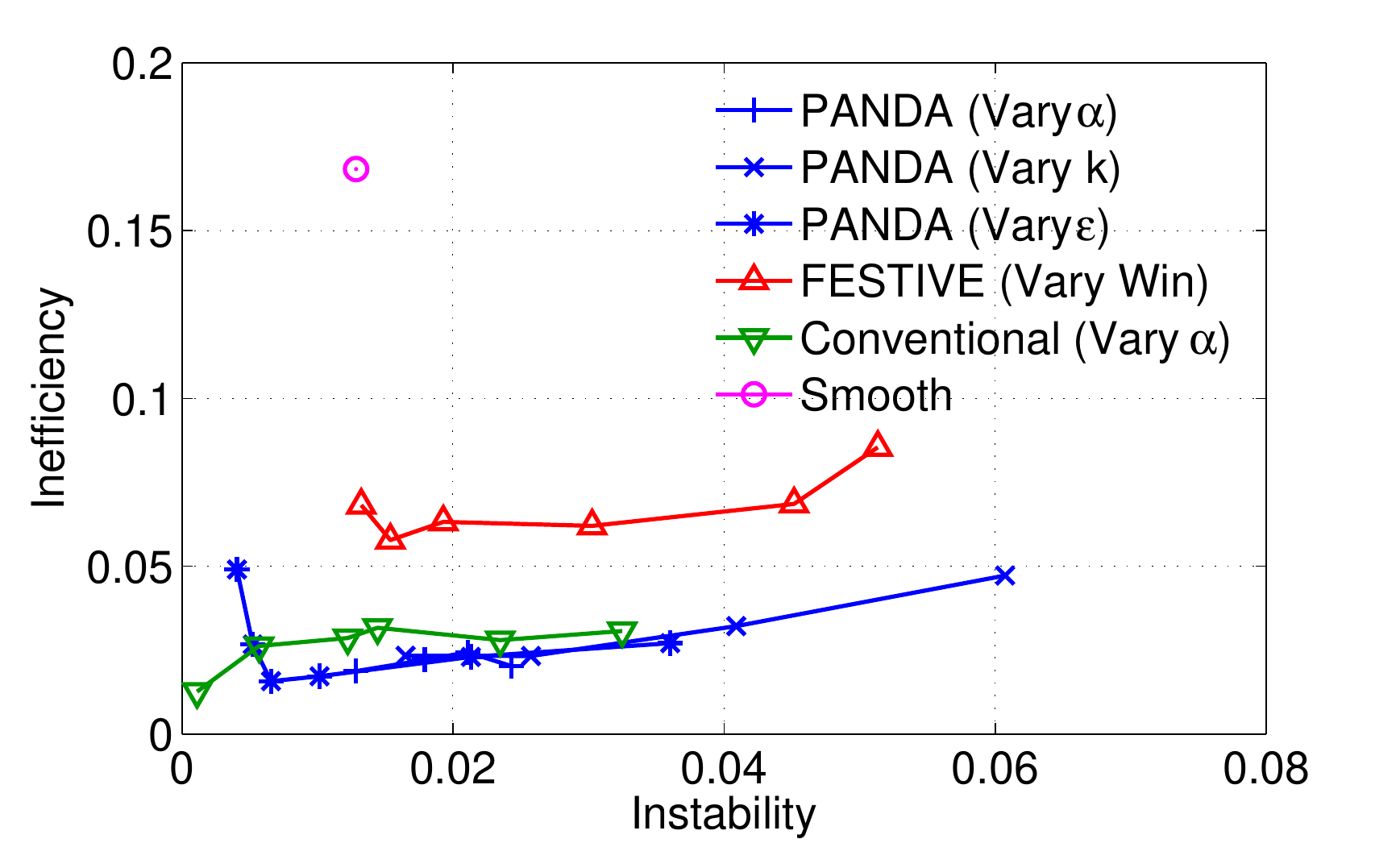}
\par\end{center}

\begin{center}
\vspace{-0.1in}
(b) 
\par\end{center}%
\end{minipage} %
\begin{minipage}[t]{0.66\columnwidth}%
\begin{center}
\includegraphics[scale=0.35]{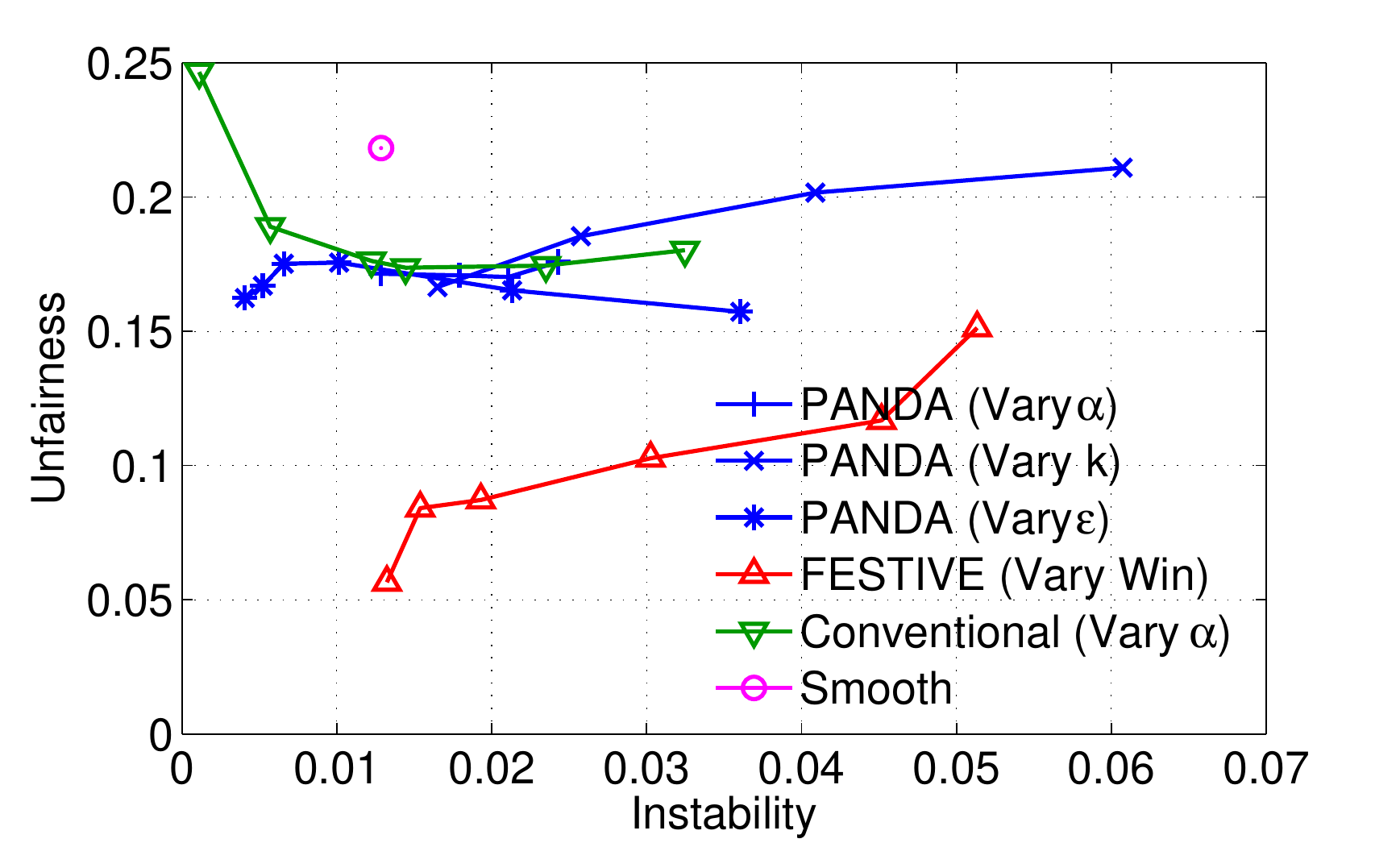}
\par\end{center}

\begin{center}
\vspace{-0.1in}
(c) 
\par\end{center}%
\end{minipage} 
\par\end{centering}

\vspace{-0.01in}

\caption{The impact of varying instability on buffer undershoot, inefficiency
and unfairness for PANDA and other benchmark players. }

\label{Flo:tradeoff} \vspace{-0.05in}
\end{figure*}

\subsection{Performance Tradeoffs}

It would not be legitimate to discuss a single metric without minding
its impact on other metrics. In this section, we examine the performance
tradeoffs among the four metrics of interest. We designed an experimental
process under which we can measure all four metrics in a single run.
For each run, five players (of the same type) compete at a bottleneck
link. The link bandwidth stays at 10 Mbps from 0 seconds to 400 seconds,
drops to 2.5 Mbps at 400 seconds and stays there until 500 seconds.
We record the instability, inefficiency and unfairness averaged over
0 to 400 seconds over all players, and the buffer undershoot over
400 to 500 seconds averaged over all players. Figure \ref{Flo:tradeoff}
shows the tradeoff between stability and each of the other criteria
-- buffer undershoot, inefficiency and unfairness -- for each of the
types of player. Each data point is obtained via averaging over 10
runs, and each data point represents a different value for one of
the parameters of the corresponding algorithm, as indicated in the
Values column of Table \ref{tab:parameters}.

For the PANDA player, the three parameters that affect instability
the most are: the probing convergence rate $\kappa$, the smoothing
convergence rate $\alpha$ and the safety margin $\epsilon$. Figure
\ref{Flo:tradeoff} (a) shows that as we vary these parameters, the
tradeoff curves mostly stay flat (except for at extreme values of
these parameters), implying that the PANDA player maintains good responsiveness
as the stability is improved. A few factors contribute to this advantage
of PANDA: First, as the bandwidth estimation by probing is quite accurate,
one does not need to apply strong smoothing. Second, since after a
bandwidth drop, the video bitrate reduction is made proportional to
the TCP throughput reduction, PANDA is very agile to bandwidth drops.
On the other hand, for both the FESTIVE and the conventional players,
the buffer undershoot significantly increases as the scheme becomes
more stable. Overall, PANDA has the best tradeoff between stability
and responsiveness to bandwidth drop, outperforming the second best
conventional player by more than 75\% reduction in instability at
the same buffer undershoot level. It is worth noting that the conventional
player uses exactly the same smoothing and quantization steps as PANDA,
which implies that the gain achieved by PANDA is purely due to the
improvement in the estimating and scheduling steps. FESTIVE has the
largest buffer undershoot. We believe this is because the design of
FESTIVE has mainly concentrated on stability, efficiency and fairness,
but ignored responsiveness to bandwidth drops. As we do not have access
to the Smooth player's buffer, we do not have its buffer undershoot
measure in Figure \ref{Flo:tradeoff} (a).

Figure \ref{Flo:tradeoff} (b) shows that PANDA has the lowest inefficiency
over all as we vary its instability. The probing mechanism ensures
that the bandwidth is most efficiently utilized. As the instability
increases, the inefficiency also increases moderately. This makes
sense intuitively, as when the bitrate fluctuates, the average fetched
bitrate also decreases. The efficiency of the conventional algorithm
underperforms PANDA, but outperforms FESTIVE. Lastly, the Smooth player
has the highest inefficiency.

Lastly, Figure \ref{Flo:tradeoff} (c) shows that in terms of fairness,
FESTIVE achieves the best performance. This may be due to the randomized
scheduling strategy of FESTIVE. PANDA and the conventional players
have similar fairness; both of them outperform the Smooth player.

\subsection{Increasing Number of Players}

\begin{figure*}
\begin{centering}
\begin{minipage}[t]{0.66\columnwidth}%
\begin{center}
\includegraphics[scale=0.35]{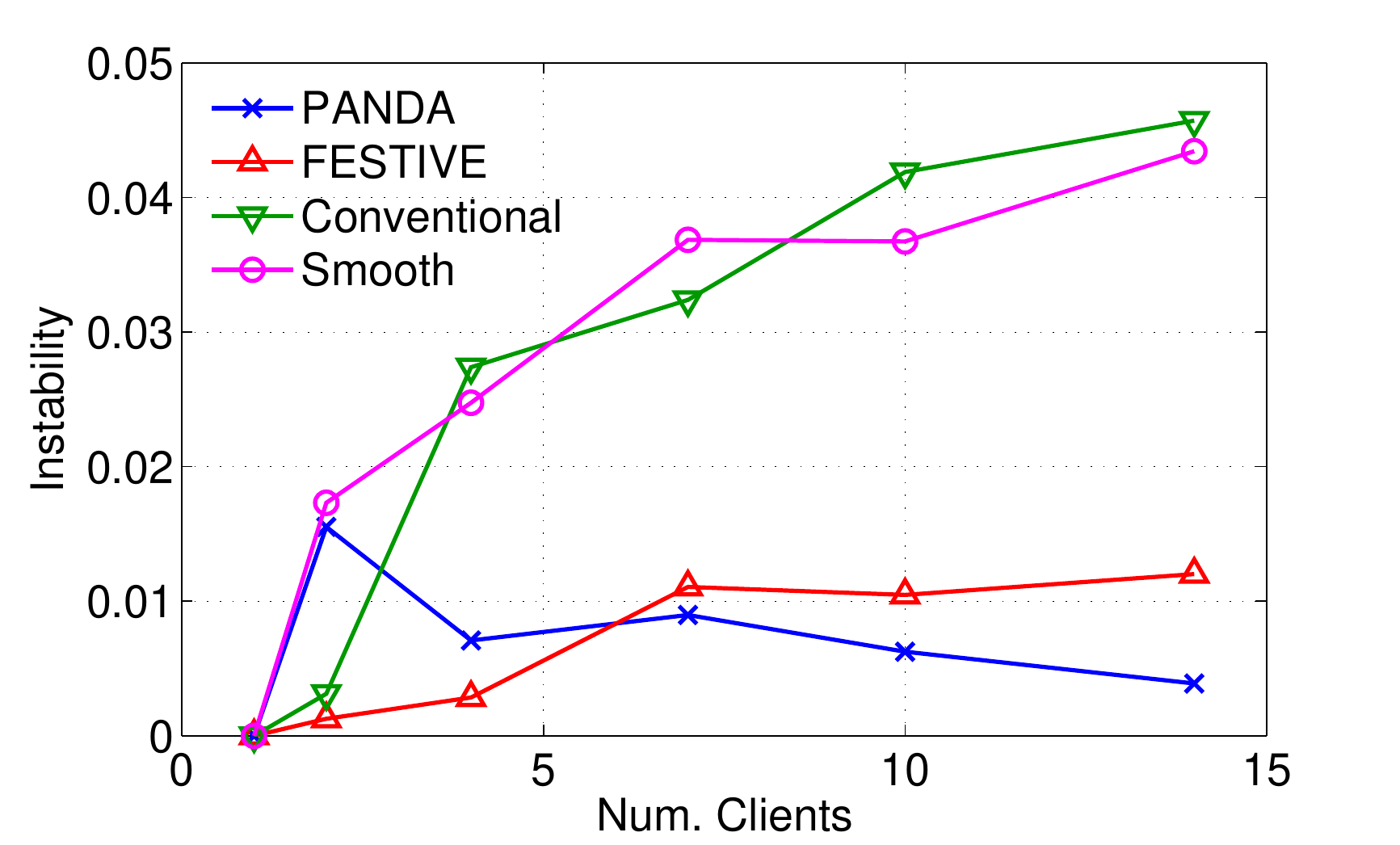}
\par\end{center}

\begin{center}
\vspace{-0.1in}
(a) 
\par\end{center}%
\end{minipage} %
\begin{minipage}[t]{0.66\columnwidth}%
\begin{center}
\includegraphics[scale=0.35]{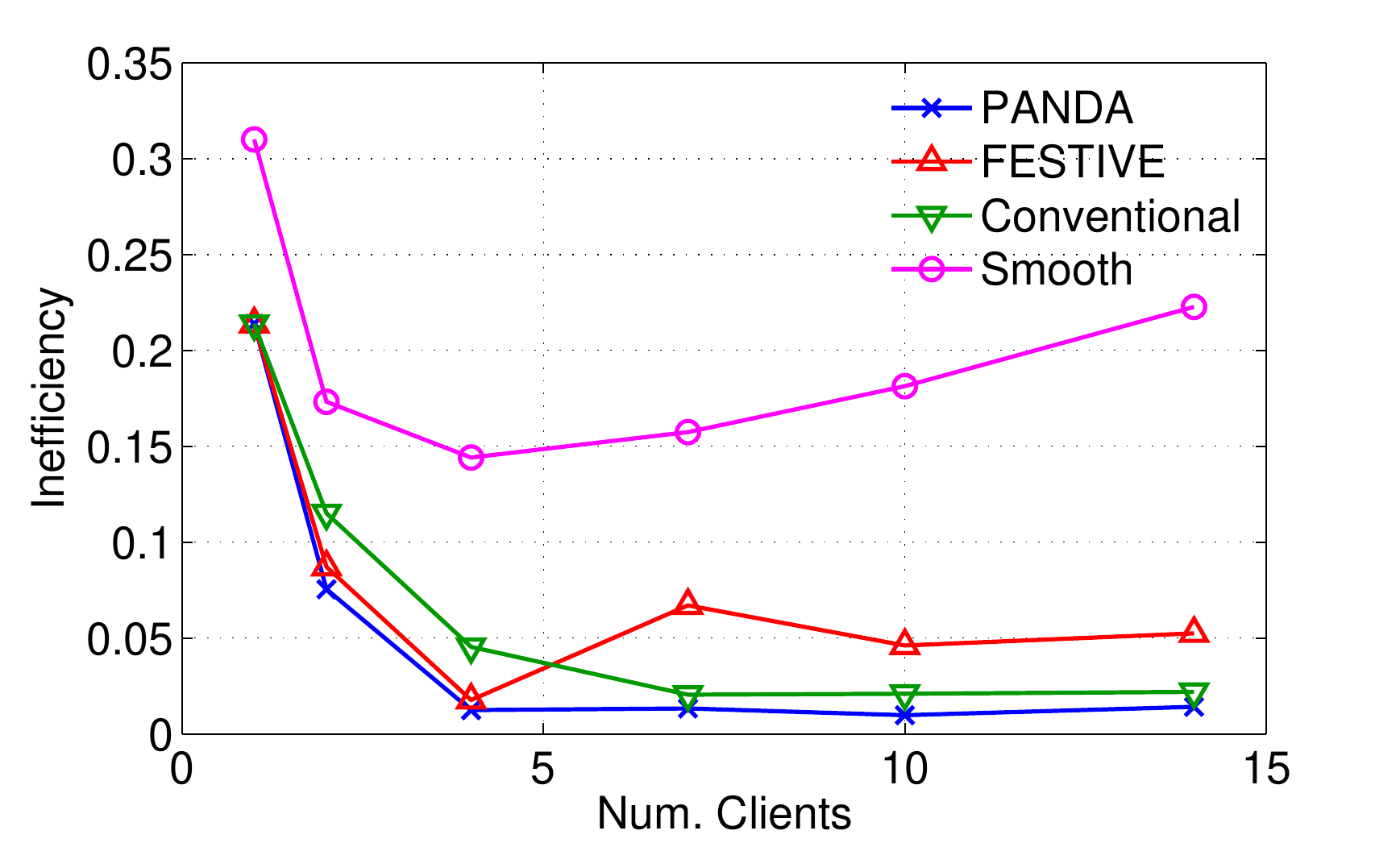}
\par\end{center}

\begin{center}
\vspace{-0.1in}
(b) 
\par\end{center}%
\end{minipage} %
\begin{minipage}[t]{0.66\columnwidth}%
\begin{center}
\includegraphics[scale=0.35]{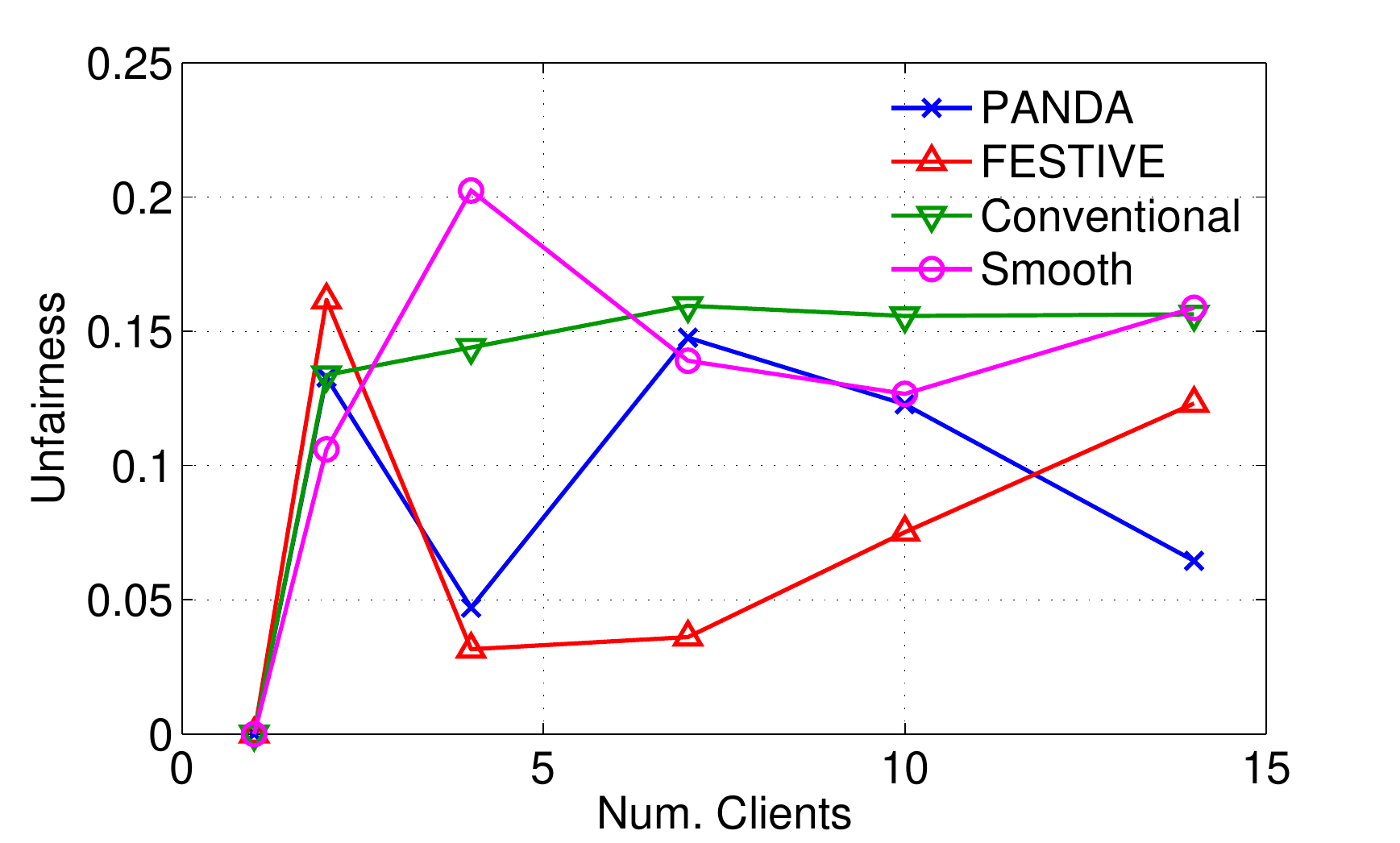}
\par\end{center}

\begin{center}
\vspace{-0.1in}
(c) 
\par\end{center}%
\end{minipage} 
\par\end{centering}

\vspace{-0.01in}

\caption{Instability, inefficiency and unfairness as the number of clients
increases. The link bandwidth is fixed at 10 Mbps.}

\label{Flo:fixbw} \vspace{-0.05in}
\end{figure*}

\begin{figure*}
\begin{centering}
\begin{minipage}[t]{0.66\columnwidth}%
\begin{center}
\includegraphics[scale=0.35]{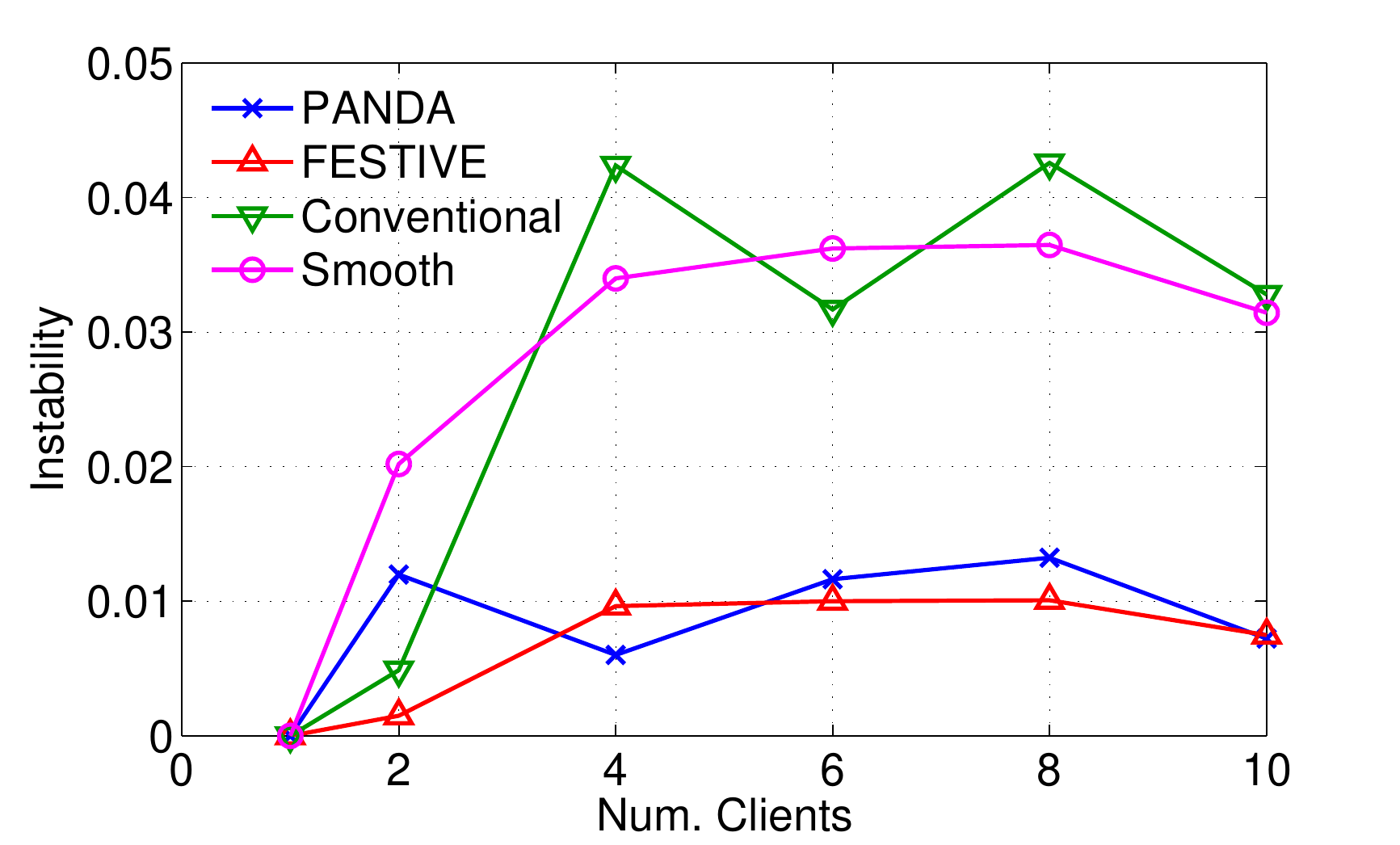}
\par\end{center}

\begin{center}
\vspace{-0.1in}
(a) 
\par\end{center}%
\end{minipage} %
\begin{minipage}[t]{0.66\columnwidth}%
\begin{center}
\includegraphics[scale=0.35]{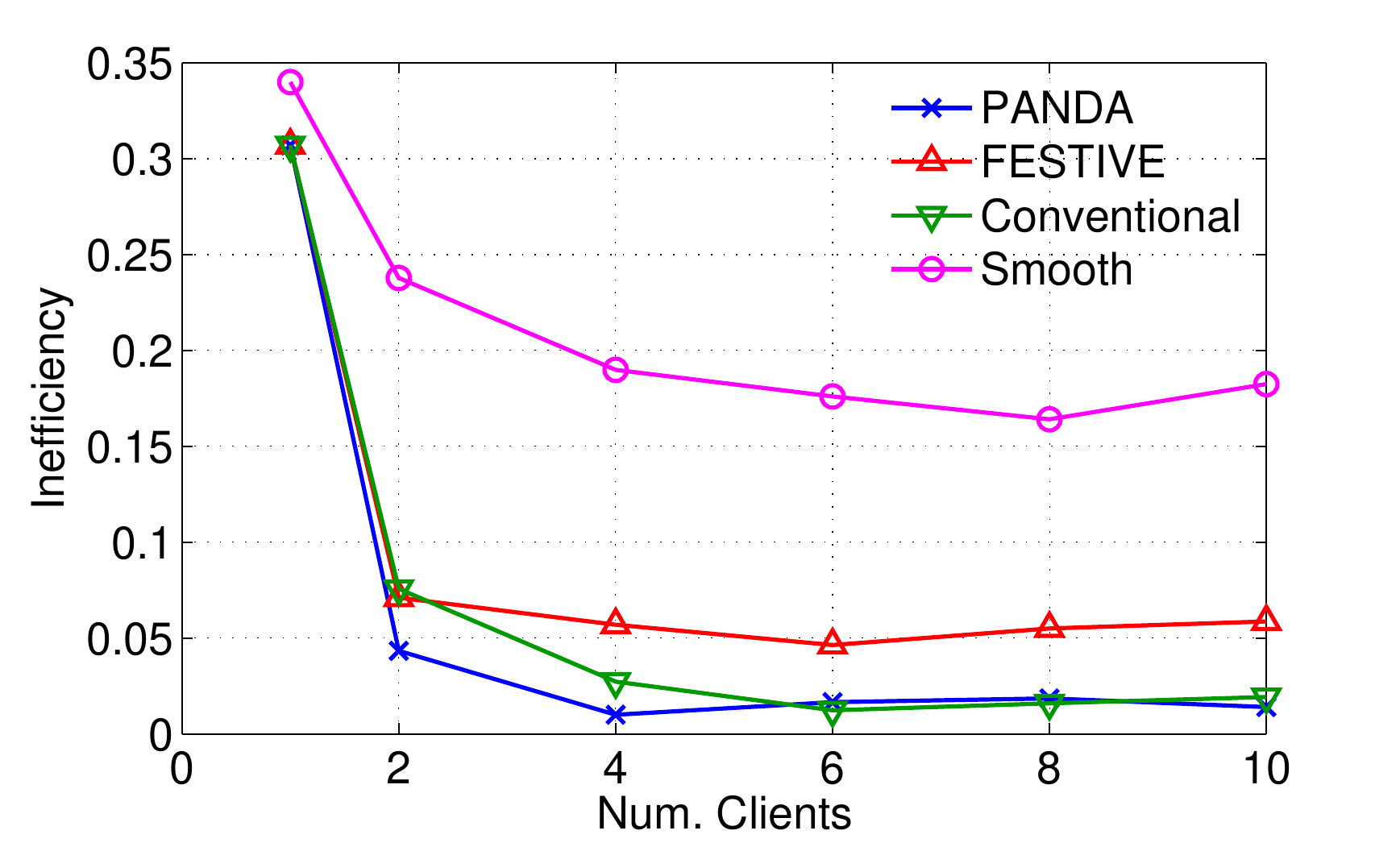}
\par\end{center}

\begin{center}
\vspace{-0.1in}
(b) 
\par\end{center}%
\end{minipage} %
\begin{minipage}[t]{0.66\columnwidth}%
\begin{center}
\includegraphics[scale=0.35]{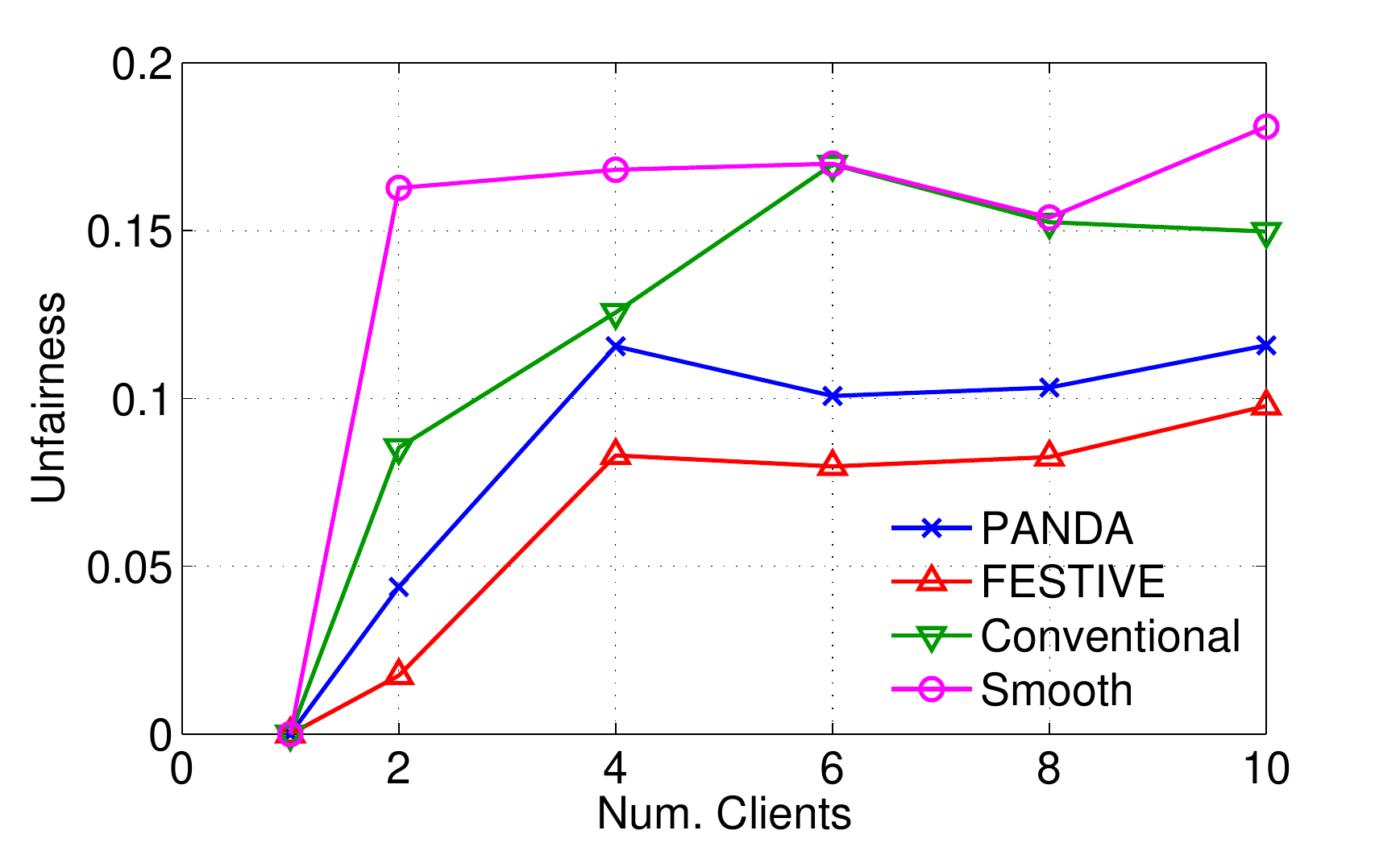}
\par\end{center}

\begin{center}
\vspace{-0.1in}
(c) 
\par\end{center}%
\end{minipage} 
\par\end{centering}

\vspace{-0.01in}

\caption{Instability, inefficiency and unfairness as the number of clients
increases. The link bandwidth increases with the number of players,
with the bandwidth-player ratio fixed at 1 Mbps/player.}

\label{Flo:varybw} \vspace{-0.05in}
\end{figure*}

In this section, we focus on the question of how the number of players
affects instability, inefficiency and unfairness at steady state.
Two scenarios are of interest: i) we increase the number of players
while fixing the link bandwidth at 10 Mbps, and ii) we increase the
number of players while varying the bandwidth such that the bandwidth/player
ratio is fixed at 1 Mbps/player. Figure \ref{Flo:fixbw} and Figure
\ref{Flo:varybw} report results for these two cases, respectively.
Each data point is obtained by averaging over 10 runs.

Refer to Figure \ref{Flo:fixbw} (a). In the single-player case, all
four schemes are able to maintain their fetched video bitrate at a
constant level, resulting in zero instability. As the number of players
increases, the instability of the conventional player and the Smooth
player both increase quickly in a highly consistent way. We speculate
that they have very similar underlying structure. The FESTIVE player
is able to maintain its stability at a lower level, due to the strong
smoothing effect (smoothing window at 20 samples by default), but
the instability still grows with the number of players, likely due
to the bandwidth overestimation effect. The PANDA player exhibits
a rather different behavior: at two players it has the highest instability,
then the instability starts to drop as the number of players increases.
Investigating into the experimental traces reveals that this is related
to the specific bitrate levels selected. More importantly, via probing,
the PANDA player is immune to the symptoms of the bandwidth overestimation,
thus it is able to maintain its stability as the number of clients
increases. The case of varying bandwidth in Figure \ref{Flo:varybw}
(a) exhibits behavior fairly consistent with Figure \ref{Flo:fixbw}
(a), with PANDA and FESTIVE exhibiting much less instability compared
to the Smooth and the conventional players.

Figure \ref{Flo:fixbw} (b) and Figure \ref{Flo:varybw} (b) for the
inefficiency metric both show that PANDA consistently has the best
performance as the number of players grow. The conventional player
and FESTIVE have similar performance, both outperforming the Smooth
player by a great margin. We speculate that the Smooth player has
some large bitrate safety margin by design, with the purpose of giving
cross traffic more breathing room.

Lastly, let us look at fairness. Refer to Figure \ref{Flo:fixbw}
(a). We have found that when the overall bandwidth is fixed, the unfairness
measure has high dependence on the specific bitrate levels chosen,
especially when the number of players is small. For example, at two
players, when the fair-share bandwidth is 5 Mbps, the two PANDA players
end up in steady state with 5.3 Mbps and 3.7 Mbps, resulting in a
high unfairness score. At three players, when the fair-share bandwidth
is 3.3 Mbps, the three PANDA players each end up with 3.7, 3.7 and
2.5 Mbps for a long period of time, resulting in a lower unfairness
score. FESTIVE exhibits lowest unfairness overall, which is consistent
with the results obtained in Figure \ref{Flo:tradeoff} (c). In the
varying-bandwidth case in Figure \ref{Flo:varybw} (c), The unfairness
ranking is fairly consistent as the number of players grow: FESTIVE,
PANDA, the conventional, and Smooth.

\subsection{Competing Mixed Players}

\begin{figure}
\begin{centering}
\hspace{-0.40in}%
\begin{minipage}[t]{1\columnwidth}%
\begin{center}
\includegraphics[scale=0.36]{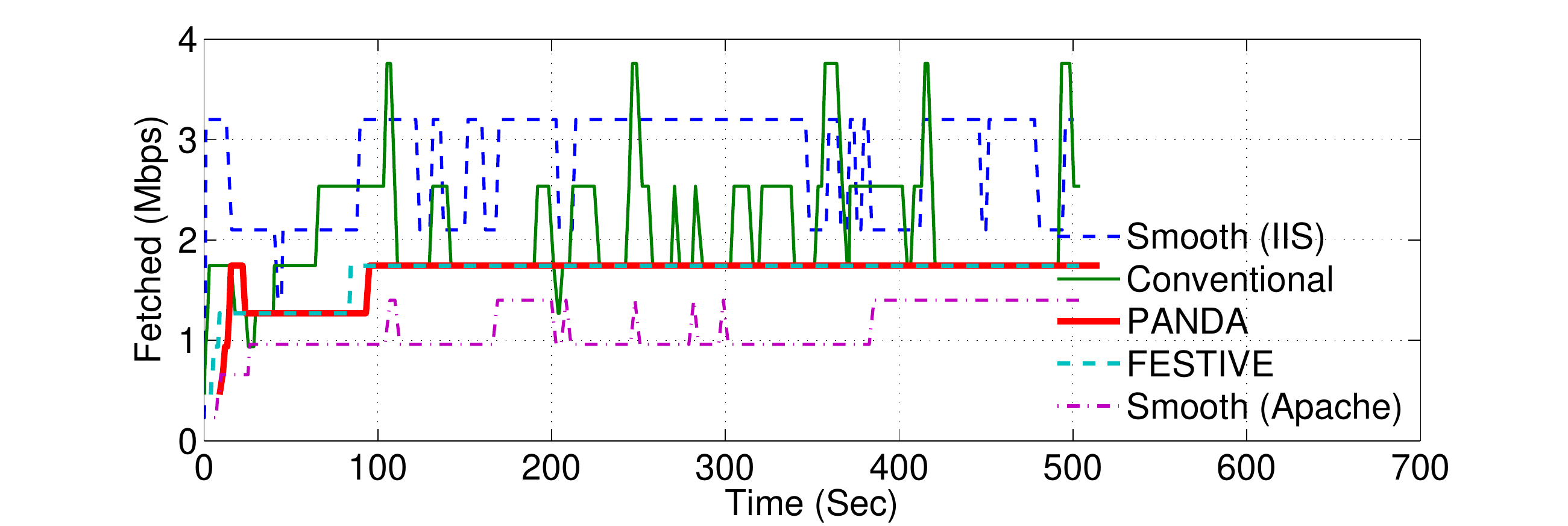}
\par\end{center}%
\end{minipage} 
\par\end{centering}

\vspace{-0.01in}\caption{PANDA, Smooth (w/ IIS), Smooth (w/ Apache), FESTIVE and the conventional
players compete at a bottleneck link of 10 Mbps.}

\label{Flo:competing} \vspace{-0.05in}
\end{figure}

One important question to ask is how PANDA will behave in the presence
of different type of players? If it behaves too conservatively and
cannot grab enough bandwidth, then the deployment of PANDA will not
be successful. To answer this question, we take the four types of
players of interest and have them compete on a 10-Mbps link. For the
Smooth player, we test it with both a Microsoft IIS server running
on Windows 7, and an Apache HTTP server running on Ubuntu 10.04. A
single trace of the fetched bitrates for 500 seconds is shown in Figure
\ref{Flo:competing}. The plot shows that the Smooth player's ability
to grab the bandwidth highly depends on the server it streams from.
Using the IIS server, which runs on Windows 7 with an aggressive TCP,
it is able to fetch video bitrates over 3 Mbps. With the Apache HTTP
server, which uses Ubuntu 10.04's conservative TCP, the fetched bitrates
are about 1 Mbps. The conventional, PANDA and FESTIVE players all
run on the same TCP (Red Hat 6), so their differences are due to their
adaptation algorithms. Due to bandwidth overestimation, the conventional
player aggressively fetches high bitrates, but the fetched bitrates
fluctuate. Both PANDA and FESTIVE are able to maintain a stable fetched
bitrate at about the fair-share level of 2 Mbps.

\subsection{Summary of Performance Results}
\begin{itemize}
\item The PANDA player has the best stability-responsiveness tradeoff, outperforming
the second best conventional player by 75\% reduction in instability.
PANDA also has the best bandwidth utilization.
\item The FESTIVE player has been tuned to yield high stability, high efficiency
and good fairness. However, it underperforms other players in responsiveness
to bandwidth drops.
\item The conventional player yields good efficiency, but lacks in stability,
responsiveness to bandwidth drops and fairness.
\item The Smooth player underperforms in efficiency, stability and fairness.
When competing against other players, its ability to grab bandwidth
is a consequence of the aggressiveness of the underlying TCP stack.
\end{itemize}

\section{Related Work\label{sec:related} }

\emph{AIMD Principle}: The design of the probing mechanism in PANDA
shares similarity with Jacobson's AIMD principle for TCP congestion
control \cite{Jacobson1988}. Kelly's framework on network rate control
\cite{kelly98} provides a theoretical justification for the AIMD
principle, and proves its stability in the general network setup.

\emph{HAS Measurement Studies}: Various research efforts have focused
on understanding the behavior of several commercially deployed HAS
systems. One such example is \cite{akhashabi12SPIC}, where the authors
characterize and evaluate HTTP streaming players such as Microsoft
Smooth Streaming, Netflix, and Adobe OSMF via experiments in controlled
settings. The first measurement study to consider HAS streaming in
the multi-client scenarios is \cite{Akhshabi:NOSSDAV12}. The authors
identify the root cause of the player's rate oscillation problem as
the existence of on-off patterns in HAS traffic. In \cite{Huang:IMC12},
the authors measure behavior of commercial video streaming services,
i.e., Hulu, Netflix, and Vudu, when competing with other long-lived
TCP flows. The results revealed that inaccurate estimations can trigger
a feedback loop leading to undesirably low-quality video. 

\emph{Existing HAS Designs}: To improve the performance of adaptive
HTTP streaming, several rate adaptation algorithms \cite{Liu:MMSys11,Tian:CoNext12,Zhou:VCIP12,Miller:PV12,Liu:SPIC12}
have been proposed, which, in general, fit into the four-step model
discussed in Section \ref{sub:Four-Step-Model}. In \cite{Jarnikov:SPIC11},
a sophisticated Markov Decision Process (MDP) is employed to compute
a set of optimal client strategies in order to maximize viewing quality.
The MDP requires the knowledge of network conditions and video content
statistics, which may not be readily available. Control-theoretical
approaches, including use of a PID controller, are also considered
by several works \cite{DeCicco:MMSys11,Tian:CoNext12,Zhou:VCIP12}.
A PID controller with appropriate parameter choice can improve streaming
performance. Server-side mechanisms are also advocated by some works
\cite{Houdaille:MMsys12,Akhshabi:NOSSDAV13}. Two designs have been
considered to address the multi-client issues: in \cite{Akhshabi:NOSSDAV13},
a rate-shaping approach aiming to eliminate the off-intervals, and
in \cite{Jiang:CoNext12}, a client rate adaptation algorithm design
implementing a combination of randomization, stateful rate selection
and harmonic mean based averaging.

\section{Conclusions\label{sec:Conclusions}}

This paper identifies an emerging issue for HTTP adaptive streaming,
which is expected to become the predominant form of the Internet traffic,
and lays out a solution direction that can effectively address this
issue. Our main contributions in this paper can be summarized as follows:
\begin{itemize}
\item We have identified the bandwidth cliff effect as the root cause of
the bitrate oscillation phenomenon and revealed the fundamental limitation
of the conventional reactive measurement based rate adaptation algorithms.
\item We have envisioned a general probe-and-adapt principle to directly
address the root cause of the problems, and designed and implemented
PANDA, a client-based rate adaptation algorithm, as an embodiment
of this principle.
\item We have proposed a generic four-step model for an HAS rate adaptation
algorithm, based on which we have fairly compared the proposed approach
with the conventional approach.
\end{itemize}
The probe-and-adapt approach and our PANDA realization thereof achieve
significant improvements in stability of HAS systems at no cost in
responsiveness. Given this framework, we plan to explore further improvements
in our future work.

{\scriptsize

\bibliographystyle{plain}
\bibliography{leeoz,ctech2012}

\begin{thebibliography}{10}

\bibitem{move07}
History of {Move Networks}.
\newblock Available online: http://www.movenetworks.com/history.html.

\bibitem{akhashabi12SPIC}
S.~Akhshabi, S.~Narayanaswamy, Ali~C. Begen, and C.~Dovrolis.
\newblock An experimental evaluation of rate-adaptive video players over
  {HTTP}.
\newblock {\em Signal Processing: Image Communication}, 27:271--287, 2012.

\bibitem{Akhshabi:NOSSDAV12}
Saamer Akhshabi, Lakshmi Anantakrishnan, Constantine Dovrolis, and Ali~C.
  Begen.
\newblock What happens when {HTTP} adaptive streaming players compete for
  bandwidth.
\newblock In {\em Proc. ACM Workshop on Network and Operating System Support
  for Digital Audio and Video (NOSSDAV'12)}, Toronto, Ontario, Canada, 2012.

\bibitem{Akhshabi:NOSSDAV13}
Saamer Akhshabi, Lakshmi Anantakrishnan, Constantine Dovrolis, and Ali~C.
  Begen.
\newblock Server-based traffic shaping for stabilizing oscillating adaptive
  streaming players.
\newblock In {\em to appear in NOSSDAV13}, 2013.

\bibitem{MSS}
{Alex Zambelli}.
\newblock {IIS Smooth Streaming Technical Overview}.
\newblock \mbox{http://tinyurl.com/smoothstreaming}.

\bibitem{DeCicco:MMSys11}
Luca~De Cicco, Saverio Mascolo, and Vittorio Palmisano.
\newblock Feedback control for adaptive live video streaming.
\newblock In {\em Proc. ACM Multimedia Systems Conference (MMSys'11)}, pages
  145--156, San Jose, CA, USA, February 2011.

\bibitem{CiscoWhitePaper}
{Cisco White Paper}.
\newblock Cisco visual networking index - forecast and methodology, 2011-2016.
\newblock {\mbox{http://tinyurl.com/ciscovni2011to2016}}.

\bibitem{zhanghui2011}
F.~Dobrian, V.~Sekar, A.~Awan, I.~Stoica, D.~Joseph, A.~Ganjam, J.~Zhan, and
  H.~Zhang.
\newblock Understanding the impact of video quality on user engagement.
\newblock In {\em Proceedings of the ACM SIGCOMM 2011 conference}, SIGCOMM '11,
  pages 362--373, New York, NY, USA, 2011. ACM.

\bibitem{Houdaille:MMsys12}
Remi Houdaille and Stephane Gouache.
\newblock Shaping http adaptive streams for a better user experience.
\newblock In {\em Proceedings of the 3rd Multimedia Systems Conference}, pages
  1--9, 2012.

\bibitem{Huang:IMC12}
Te-Yuan Huang, Nikhil Handigol, Brandon Heller, Nick McKeown, and Ramesh
  Johari.
\newblock Confused, timid, and unstable: Picking a video streaming rate is
  hard.
\newblock In {\em Proceedings of the 2012 ACM conference on Internet
  measurement conference}, 2012.

\bibitem{Jacobson1988}
V.~Jacobson.
\newblock Congestion avoidance and control.
\newblock In {\em Symposium proceedings on Communications architectures and
  protocols}, SIGCOMM '88, pages 314--329, New York, NY, USA, 1988. ACM.

\bibitem{Jarnikov:SPIC11}
Dmitri Jarnikov and Tanir Ozcelebi.
\newblock {Client Intelligence for Adaptive Streaming Solutions}.
\newblock {\em {EURASIP Journal on Signal Processing: Image Communication,
  Special Issue on Advances in IPTV Technologies}}, 26(7):378--389, August
  2011.

\bibitem{Jiang:CoNext12}
Junchen Jiang, Vyas Sekar, and Hui Zhang.
\newblock Improving fairness, efficiency, and stability in http-based adaptive
  video streaming with festive.
\newblock In {\em Proceedings of the 8th international conference on Emerging
  networking experiments and technologies (CoNEXT)}, 2012.

\bibitem{kelly98}
F.~P. Kelly, A.~K. Maulloo, and D.~K.~H. Tan.
\newblock Rate control for communication networks: Shadow prices, proportional
  fairness and stability.
\newblock {\em The Journal of the Operational Research Society},
  49(3):237--252, 1998.

\bibitem{Liu:MMSys11}
Chenghao Liu, Imed Bouazizi, and Moncef Gabbouj.
\newblock {Rate Adaptation for Adaptiv HTTP streaming}.
\newblock In {\em Proc. ACM Multimedia Systems Conference (MMSys'11)}, pages
  169--174, San Jose, CA, USA, February 2011.

\bibitem{Liu:SPIC12}
Chenghao Liu, Imed Bouazizi, Miska~M. Hannuksela, and Moncef Gabbouj.
\newblock {Rate adaptation for dynamic adaptive streaming over HTTP in content
  distribution network}.
\newblock {\em Signal Processing: Image Communication}, 27(4):288--311, April
  2012.

\bibitem{Miller:PV12}
Konstantin Miller, Emanuele Quacchio, Gianluca Gennari, and Adam Wolisz.
\newblock Adaptation algorithm for adaptive streaming over http.
\newblock In {\em Proceedings of 2012 IEEE 19th International Packet Video
  Workshop}, pages 173--178, 2012.

\bibitem{Mok:WMUST2011}
R.K.P. Mok, E.W.W. Chan, X.~Luo, and R.K.C. Chang.
\newblock Inferring the {QoE} of {HTTP} video streaming from user-viewing
  activities.
\newblock In {\em Proc. SIGCOMM W-MUST}, 2011.

\bibitem{Tian:CoNext12}
Guibin Tian and Yong Liu.
\newblock Towards agile and smooth video adaptation in dynamic http streaming.
\newblock In {\em Proceedings of the 8th international conference on Emerging
  networking experiments and technologies (CoNEXT)}, pages 109--120, 2012.

\bibitem{Zhou:VCIP12}
Chao Zhou, Xinggong Zhang, Longshe Huo, and Zongming Guo.
\newblock A control-theoretic approach to rate adaptation for dynamic http
  streaming.
\newblock In {\em Proceedings of Visual Communications and Image Processing
  (VCIP)}, pages 1--6, 2012.

\end{thebibliography}

}

\appendices{}

\section{Bandwidth Cliff Effect: Theoretical Analysis\label{sec:Bandwidth-Cliff-Effect:}}

\subsection{Problem Formulation\label{sec:Problem-Formulation}}

Consider that $K$ clients share a bottleneck link of capacity $C$.
The streaming process of each client consists of discrete downloading
steps $n=1,2,...$, where during each step one video segment is downloaded. 

\vspace{0.05in}

\emph{Fixed requested video bitrate.} As we are interested in how
the measured TCP throughput causes HAS clients to shift their video
bitrates requested, we analyze the stage of the dynamics where the
rate shift \emph{has not occurred} \emph{yet}. Thus, in our model,
we assume that each HAS client does not change its requested video
bitrate over the time interval of analysis. For $k=1,...K$, each
$k$-th client requests a video segment of fixed size $r_{k}\cdot\tau$
at each downloading step, where $r_{k}\in\mathcal{R}$ is the video
bitrate selected.

\vspace{0.05in}

\emph{Segment requesting time and downloading duration. }Denote by
$R_{k}(t)$ the instantaneous data downloading rate of the $k$-th
client at time $t$ (note that $R_{k}$ and $r_{k}$ are different).
Denote by $t_{k}[n]$ the time that the $k$-th client requests its
$n$-th segment (which, for simplicity, is also assumed to be the
time that the downloading starts). For each client, assume that the
requesting time of the first segment $t_{k}[1]$ is given. For $n\geq1$,
the requesting time of the next segment is determined by:
\begin{equation}
t_{k}[n+1]=t_{k}[n]+\max(\tau,\tilde{T_{k}}[n]),\label{eq:start_time1}
\end{equation}
where $\tilde{T_{k}}[n]$ is the actual duration of the $k$-th client
downloading its $n$-th segment. This is a reasonable assumption where
the buffer level of a HAS client has reached the maximum level. The
actual duration of downloading, $\tilde{T_{k}}[n]$, can be related
to $r_{k}$ by
\begin{equation}
\int_{t=t_{k}[n]}^{t_{k}[n]+\tilde{T_{k}}[n]}R_{k}(t)\cdot dt=r_{k}\cdot\tau.\label{eq:duration1}
\end{equation}
The \emph{TCP throughput} measured by the $k$-th client for downloading
the $n$-th segment, is defined as $\tilde{x}_{k}[n]:=\frac{r_{k}\cdot\tau}{\tilde{T_{k}}[n]}$.
In a conventional HAS algorithm, the TCP throughput is used as an
estimator of a client's fair-share bandwidth, which, ideally, is equal
to $\frac{C}{K}$.

\vspace{0.05in}

\emph{Idealized TCP behavior.} We assume that the network obeys a
simplified bandwidth sharing model, where we do not consider the effects
such as TCP slow-start restart and heterogenous RTTs.%
\footnote{It is trivial to extend the analysis to the case of heterogenous RTTs.%
} At any moment $t\geq0$ when there are $A(t)\geq1$ active TCP flows
sharing the bottleneck link, each active flow will receive a fair-share
data rate of $R_{k}(t)=\frac{C}{A(t)}$ instantaneously, and the total
traffic rate in the link is $C$; at any moment when there is no active
TCP flows, or $A(t)=0$, the total traffic rate in the link is $0$.
For this case, we have the following definition:
\begin{defn}
A \emph{gap} is an interval within which the total traffic rate of
all clients is $0$.
\end{defn}
\vspace{0.05in}

\emph{Client initial state. }We assume that each client may have some
arbitrary initial state, including:
\begin{itemize}
\item Time of requesting the first segment $t_{k}[1]$, $k=1,...,K$, as
forementioned.
\item Initial data downloading rate, i.e., it may be that $R_{k}(t)>0$
for $t<t_{k}[1]$, $k=1,...,K$, where the rate may be due to requesting
a segment earlier than the first segment being considered. In practice,
this may correspond to the cases where the clients have already started
downloading but may be in a different state, before the first segment
being considered (e.g., link is oversubscribed before it becomes undersubscribed).
\end{itemize}

\subsection{Link Undersubscription}

The link is \emph{undersubscribed} by the $K$ HAS clients if the
sum of the requested video bitrates is less than the link capacity,
i.e., $\sum_{k=1}^{K}r_{k}<C$. We would like to show that even the
\emph{slightest} undersubscription of the link would lead to convergence
into a state where each client has a TCP throughput greater than its
fair-share bandwidth $\frac{C}{K}$.

To begin with, we prove a set of lemmas. We first show that any two
adjacent requesting times $t_{k}[n]$ and $t_{k}[n+1]$ are spaced
by exactly $\tau$ if there exists a gap between them.
\begin{lem}
\label{lem:offinterval1}$t_{k}[n+1]=t_{k}[n]+\tau$ if there exists
a gap $[t^{-},t^{+})$ with $t_{k}[n]\leq t^{-}$ and $t^{+}\leq t_{k}[n+1]$. \end{lem}
\begin{IEEEproof}
By (\ref{eq:start_time1}), the only case that $t_{k}[n+1]\neq t_{k}[n]+\tau$
is when $\tilde{T_{k}}[n]>\tau$. But this cannot hold, since otherwise
there cannot be a gap $[t^{-},t^{+})$ with $t_{k}[n]\leq t^{-}$
and $t^{+}\leq t_{k}[n+1]$.
\end{IEEEproof}
The rest of the lemmas in this section make the assumption of $\sum_{k=1}^{K}r_{k}<C$.
First, we show that at least one gap must emerge after some time.
\begin{lem}
\label{lem:undersubscribed1}There exists a gap $[t^{-},t^{+})$,
where $ $$t^{-}>\max_{k}t_{k}[1]$.\end{lem}
\begin{IEEEproof}
By (\ref{eq:start_time1}), within $[t,t+\tau)$ for any $t$, each
client can download at most one segment, or data of size at most $r_{k}\cdot\tau$.
Consider within an interval $[\max_{k}t_{k[1]},\max_{k}t_{k[1]}+m\cdot\tau)$
for some $m\in\mathcal{N}$. The maximum size of the data that can
be downloaded by the $K$ clients is $B_{res}+m\cdot\sum_{k=1}^{K}r_{k}\tau$,
where $B_{res}$ is the total size of the residue data from segments
requested prior to $\max_{k}t_{k}[1]$, including those prior to the
first segments being considered, as discussed in Section \ref{sec:Problem-Formulation},
\emph{client initial state}. By the idealized TCP behavior, at any
instant, the total traffic rate can be either $C$ or $0$. If zero
total traffic rate does not happen, the total downloaded data size
within the interval being considerd is $mC\tau$. Therefore, a sufficient
condition for a gap $[t^{-},t^{+})$ to occur is to have $t^{-}>\max_{k}t_{k[1]}+m\cdot\tau$,
where $mC\tau>B_{res}+m\cdot\sum_{k=1}^{K}r_{k}\tau$, or $m=\left\lceil B_{res}/\tau\cdot(C-\sum_{k=1}^{K}r_{k})\right\rceil $.
\end{IEEEproof}
The next lemma shows that within an interval of duration $\tau$ immediately
following this gap, each client must request one and only one segment.
\begin{lem}
Within the interval $ $$[t^{+},t^{+}+\tau)$, each client must request
one and only one segment.\end{lem}
\begin{IEEEproof}
First, we show that each client must request at least one segment.
Invoking Lemma \ref{lem:offinterval1} and the fact that $[t^{-},t^{+})$
is a gap, the request times immediately preceding and following $[t^{-},t^{+})$
must be spaced exactly by $\tau$. This can never hold if no segment
is requested within $[t^{+},t^{+}+\tau)$. 

Second, we show that each client can request at most one segment within
$[t^{+},t^{+}+\tau)$. This directly follows applying (\ref{eq:start_time1})
to any interval of duration $\tau$, similar to the proof of Lemma
\ref{lem:undersubscribed1}. 
\end{IEEEproof}
Since exactly one segment is requested by each client within $[t^{+},t^{+}+\tau)$,
for convenience, let us re-label these segments using a new index
$n'$.
\begin{lem}
\label{lem:offinterval12}$[t^{-}+\tau,t^{+}+\tau)$ is a gap.\end{lem}
\begin{IEEEproof}
First, invoking Lemma \ref{lem:offinterval1} and the fact that $[t^{-},t^{+})$
is a gap, we have $t_{k}[n'-1]=t_{k}[n']-\tau$ for $k=1,...,K$.
In other words, the starting time patterns within intervals $[t^{+}-\tau,t^{+})$
and $[t^{+},t^{+}+\tau)$ exactly repeat.

Second, consider the data traffic within $[t^{+}-\tau,t^{+})$ and
$[t^{+},t^{+}+\tau)$. The only difference is that within $[t^{+}-\tau,t^{+})$,
there might be unfinished residue data from the previous segments
whereas within $[t^{+},t^{+}+\tau)$, there is no such unfinished
residue data due to the gap $[t^{-},t^{+})$. By (\ref{eq:duration1}),
the exact starting times and the idealized TCP behavior, the downloading
completion time can only get delayed with the extra residue data,
thus we must have $ $$\tilde{T}_{k}[n'-1]\geq\tilde{T}_{k}[n']$,
$k=1,...,K$. Therefore, the data traffic within $[t^{+},t^{+}+\tau)$
must finish no later than $t^{-}+\tau$, and $[t^{-}+\tau,t^{+}+\tau)$
must also be a gap.
\end{IEEEproof}
The following theorem shows that in the case of link undersubscription,
regardless of the client initial states, the banwidth sharing among
the clients will eventually converge to a periodic pattern.
\begin{thm}
\label{thm:undersubscribed1}If $\sum_{k=1}^{K}r_{k}<C$, the data
downloading rate $R_{k}(t)$ of each client will converge to a periodic
pattern with period $\tau$, i.e., there exists some $t'\geq0$ such
that for all $t\geq t'$, $R_{k}(t+\tau)=R_{k}(t)$, $k=1,...,K$.\end{thm}
\begin{IEEEproof}
The interval $[t^{+},t^{+}+\tau)$ has no residue data from the previous
unfinished segments, because of the gap $[t^{-},t^{+})$. Using this
fact and the idealized TCP behavior, the data downloading rates $R_{k}(t)$,
$k=1,...,K$ for $t\in[t^{+},t^{+}+\tau)$ is a deterministic function
of the starting times $t_{k}[n']$, $k=1,...,K$. The same argument
applies to $R_{k}(t)$, $k=1,...,K$ for $t\in[t^{+}+\tau,t^{+}+2\tau)$
and $t_{k}[n'+1]$, $k=1,...,K$. 

By Lemma \ref{lem:offinterval1} and Lemma \ref{lem:offinterval12},
we have $t_{k}[n'+1]=t_{k}[n']+\tau$ for $k=1,...,K$. In other words,
other than the offset $\tau$, the starting time patterns within $[t^{+},t^{+}+\tau)$
and $[t^{+}+\tau,t^{+}+2\tau)$ are the same. Invoking the deterministic
function argument, we can show $R_{k}(t+\tau)=R_{k}(t)$, $k=1,...,K$
for $t\in[t^{+},t^{+}+\tau)$. Taking $t'=t^{+}$ and repeatly applying
the argument above to the intervals $[t^{+}+m\tau,t^{+}+(m+1)\tau)$,
$m=2,3,...$ complete the proof.
\end{IEEEproof}
Note that by the idealized TCP behavior, no client's TCP throughput
would overestimate its fair-share bandwdith $\frac{C}{K}$ \emph{only
if} the number of active flows $A(t)=K$ during all the active time.
After the bandwith sharing pattern converges to periodic, this happens
only when the starting times of all clients are all aligned and their
segment data size are equal. The following corollary is an immediate
consequence of Theorem \ref{thm:undersubscribed1}:
\begin{cor}
If $\sum_{k=1}^{K}r_{k}<C$, and $t_{k}[n']\neq t_{l}[n']$ for some
$k\neq l$, then for some clients the TCP throughput will converge
to a value that is greater than its fair-share bandwdith, i.e., there
exists $k'$ with $\tilde{x}_{k'}[n]>\frac{C}{K}$ for $n\geq n'$.
In particular, if $r_{1}=r_{2}=...=r_{K}$, then for all $k'=1,...,K$,
$\tilde{x}_{k'}[n]>\frac{C}{K}$ for $n\geq n'$.
\end{cor}
Note that the start times pattern $t_{k}[n']$, $k=1,...,K$ will
dictate exactly by how much the TCP throughput overestimates the fair-share
bandwidth, with the range $\frac{\tilde{x}_{k}[n]}{C/K}\in[1,K]$.

\subsection{Link Oversubscription}

We next consider the case that the link is \emph{oversubscribed} by
the $K$ HAS clients, i.e., $\sum_{k=1}^{K}r_{k}>C$. We first show
a sufficient condition under which the TCP throughput would correctly
predict the fair-share bandwidth.
\begin{thm}
\label{thm:oversubscribed2}If $r_{k}>\frac{C}{K}$ for all $k=1,...,K$,
all clients' TCP throughput converges to the fair-share bandwidth,
i.e., there exists $n'>0$ such that $\tilde{x}_{k}[n]=\frac{C}{K}$,
$k=1,...,K$ for $n\geq n'$.\end{thm}
\begin{IEEEproof}
We first want to show that there exists $n'>0$ such that $\tilde{T}_{k}[n]>\tau$
for all $k=1,...,K$ for $n\geq n'$. Assume that the opposite is
true, i.e., there exists at least a client $k'$ such that, $\tilde{T}_{k'}[n]\leq\tau$
for all $n\geq n'$. Since $r_{k'}>\frac{C}{K}$, this would hold
only if within the active intervals of client $k'$, at least another
client $k''$ must be inactive. Denote by $[t_{k''}^{-},t_{k''}^{+})$
the first such inactive interval. Consider within an interval $[t_{k''}^{-},t_{k''}^{-}+m\cdot\tau)$
for some $m\in\mathcal{N}$. By the idealistic TCP behavior, the total
number of bits that can be downloaded by the $K-1$ clients (other
than $k'$) must be $\leq(K-1)/K\cdot m\tau C$. This contradicts
the fact that $\tilde{T}_{k}[n]>\tau$, for $k\neq k'$, $n\geq n'$.
Thus, the assumption is invalid.

Then, by (\ref{eq:start_time1}), for $n\geq n'$, all clients are
active all the time. By the idealistic TCP behavior, we have $\tilde{x}_{k}[n]=\frac{C}{K}$,
$k=1,...,K$ for $n\geq n'$.
\end{IEEEproof}
By slightly extending the above argument, a sufficient and necessary
condition for the correct fair-share bandwidth estimation of \emph{all
clients} can be found.
\begin{thm}
\label{thm:oversubscribed3}If $\sum_{k=1}^{K}r_{k}>C$, all clients'
TCP throughput converges to the fair-share bandwidth, if and only
if $r_{k}\geq\frac{C}{K}$ for all $k=1,...,K$ and there exists $k'$
such that $r_{k'}>\frac{C}{K}$.
\end{thm}

\section{Analysis of PANDA\label{sec:Analysis-of-PANDA}}

We perform an equilibrium and stability analysis of PANDA. For simplicity,
we only analyze the single-client case where the system has an equilibrium
point.

\subsection{Analysis of $\hat{x}$}

Assume the measured TCP download throughput $\tilde{x}$ is equal
to the link capacity $C$. Consequently, \eqref{eq:baseline2b} can
be re-written in two scenarios (refer to Figure \ref{Flo:delay2})
as:
\begin{equation}
\frac{\hat{x}[n]-\hat{x}[n-1]}{T[n-1]}=\begin{cases}
\kappa\cdot w, & \mbox{if~}\hat{x}<C,\\
\kappa\cdot w-\kappa\cdot(\hat{x}[n-1]-C), & \mbox{otherwise}.
\end{cases}\label{eqn:newx}
\end{equation}

At equilibrium, setting \eqref{eqn:newx} to zero leads to $w=\hat{x}_{o}-C$,
hence, $\hat{x}_{o}=C+w>\tilde{x}_{o}=C$.

Considering the simplifying assumptions of $\hat{y}=\hat{x}$ (no
smoothing) and a quantizer with a margin $\Delta$ such that $r=\hat{y}-\Delta$,
we need $r_{o}=\hat{x}_{o}-\mbox{\ensuremath{\Delta}}<C$ at equilibrium.
Therefore, the quantization margin of the quantizer needs to satisfy
\[
\Delta>\hat{x}_{o}-C=w,
\]
so that the system stays on the multiplicative-decrease side of the
equation at steady state.

Note that close to equilibrium, the intervals between consecutive
segment downloads match the segment playout duration $T[n-1]=\tau$,
therefore, calculation of the estimated bandwidth $\hat{x}$ follows
the simple form of a difference equation:
\[
\hat{x}[n]=a\cdot\hat{x}[n-1]+b.
\]
The two constants are: $a=1-\kappa\cdot\tau$ and $b=\kappa\cdot(C+w)\cdot\tau$.
The sequence converges if and only if $|a|<1$. Hence, we need: $-1<1-\kappa\cdot\tau<1$,
leading to the stability criterion: 
\[
0<k<\frac{2}{\tau}.
\]

\subsection{Analysis of $\hat{T}$ and $B$}

Assume no smoothing $\hat{x}=\hat{y}$. During transient states, when
$T=\hat{T}>\tilde{T}$, update of $\hat{T}$ pushes the playout buffer
size in the right direction: $\hat{T}<\frac{r[n]\cdot\tau}{\hat{x}[n]}$
leads to a steady growth in buffer size when $B<B_{\min}$. At steady
state, $\hat{T}_{o}=\tau=\hat{T}_{o}>\tilde{T}_{o}$. It can be derived
from \eqref{eq:baseline3b} that: 
\[
B_{o}=B_{\min}+\left(1-\frac{r_{o}}{\hat{x}_{o}}\right)\cdot\frac{\tau}{\beta}
\]
where $B_{o}$ is playout buffer at equilibrium.
\end{document}